\documentclass[twocolumn,aps,prb,showpacs,superscriptaddress,
amsmath,amssymb]{revtex4}
\usepackage{graphicx}
\usepackage{color} 
\newcommand{\beq}{\begin{equation}}
\newcommand{\eneq}{\end{equation}}
\newcommand{\be}{\begin{equation}}
\newcommand{\ee}{\end{equation}}
\newcommand{\bea}{\begin{eqnarray}}
\newcommand{\eea}{\end{eqnarray}}

\begin{document}

\title{Chirality and Current-Current Correlation in Fractional Quantum Hall Systems.}
\author{G. Campagnano}
\altaffiliation{Contributed equally to this work.}
\affiliation{CNR-SPIN, Monte S.Angelo -- via Cinthia,  I-80126 Napoli, Italy}
\affiliation{Dipartimento di Fisica ``E. Pancini", Universit\`a di Napoli ``Federico II'', 
Monte S.Angelo, I-80126 Napoli, Italy}

\author{P. Lucignano}
\affiliation{CNR-SPIN, Monte S.Angelo -- via Cinthia,  I-80126 Napoli, Italy}
\affiliation{Dipartimento di Fisica ``E. Pancini", Universit\`a di Napoli ``Federico II'', 
Monte S.Angelo, I-80126 Napoli, Italy}

\author{D. Giuliano}
\altaffiliation{Contributed equally to this work.}
\affiliation{Dipartimento di Fisica, Universit\`a della Calabria and INFN, Gruppo 
Collegato di Cosenza Arcavacata di Rende, I-87036, Cosenza, Italy}

\begin{abstract}
We study current-current correlation in an electronic analog of a beam splitter realized with edge channels of a fractional quantum Hall liquid at Laughlin filling fractions. In analogy with  the known result for chiral electrons\cite{Buettiker:1992}, if the currents are measured at points located after the beam splitter, we find that  the zero frequency equilibrium correlation vanishes due to the chiral propagation along the edge channels. Furthermore, we show that the current-current correlation, normalized to the tunneling current, exhibits clear signatures of the Laughlin quasi-particles' fractional statistics. 
\end{abstract}

\maketitle

\section{Introduction}

With the theoretical explanation of the Fractional Quantum Hall (FQH) effect at 
filling $\nu = 1 / m$, with $m$ odd \cite{tsui}, 
Laughlin made the remarkable prediction that the elementary charged excitation of
a FQH system is a quasiparticle carrying fractional charge $q = \nu e$. \cite{laughlin}
Moreover, Laughlin's quasiparticles  (LQP) were predicted to carry 
fractional statistics, as well, that is, on exchanging two of them with each other, 
the relative wavefunction must acquire a statistical phase $\theta\ne\{ \pi,2\pi\}$, 
with $\theta = \pi$ corresponding to fermions, $\theta = 2 \pi$ corresponding 
to bosons. As a result, they behave as Abelian anyons with fractional charge $q$.\cite{Arovas:1984} 

Even richer structures are possible in the case of non-Abelian anyons---the braiding of one 
quasi-particle by another one will result in the system to be sent into a 
different quantum state and not only in the relative wavefunction to acquire a 
statistical phase\cite{Najak:2008}.  

A FQH system is fully gapped in the bulk, with gapless branches 
of chiral excitations at its edges, supporting current flow across the 
sample. The elementary charge carrier  is the boundary 
analog of a LQP and, therefore, it carries a fractional charge $q$, as well \cite{wenbook}. 
Because of such a correspondence, it was possible to experimentally estabilish the fractional charge of 
LQPs  by means of shot-noise measurements on a FQH-bar \cite{de-Picciotto:1997,Saminadayar:1997}.
Nevertheless,  a direct observation of their fractional statistics is still the subject of 
ongoing experimental efforts\cite{Camino:2005,Willett:2013}. 

Correlation measurements of light intensities in optics\cite{Hanbury:1954,Hanbury:1955} 
and electrical currents in solid-state physics\cite{Oliver:1999}  have provided an 
important tool to investigate the difference between the two ''classical'' statistics of 
quantum elementary particles: bosonic and fermionic. In the pursue of evidence for fractional statistics 
in FQH systems, a number of works have been putting forward the use of solid-state analogs  of 
Fabry-Perot\cite{Chamon:1997,Stern:2006,Bonderson:2006,Bonderson:2006b,Kim:2008,Bishara:2009,Halperin:2011}, 
Mach-Zehnder\cite{Feldman:2006,Feldman:2007,Ponomarenko:2007,Ponomarenko:2009,Bonderson:2008,Law:2008,Levkivskyi:2009,Ponomarenko:2010,Wang:2010,Levkivskyi:2012,Ganeshan:2012,Yang:2015} and 
more elaborated Hanbury Brown and Twiss
\cite{Safi:2001,Vishveshwara:2003,Kim:2005,Kim:2006,Campagnano:2012,Campagnano:2013} 
interferometers to address the statistical properties  of fractional quantum Hall anyons.

\begin{figure}[h]
\includegraphics[width=0.75\linewidth]{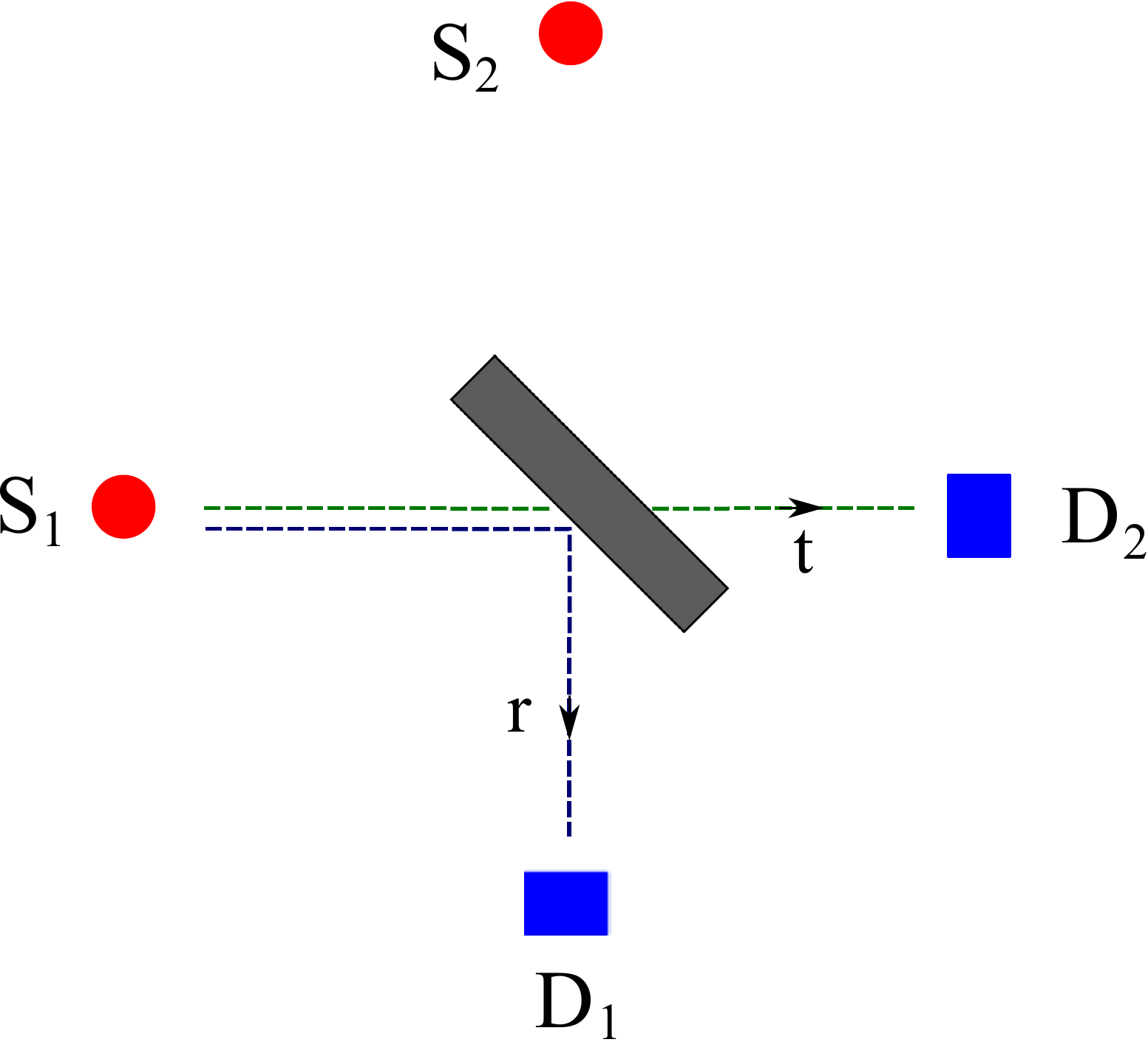}
\caption{In Figure we represent a scheme of a beam splitter. Particles emitted from source $S_1$ can be transmitted to detector $D_2$ with transmission amplitude $t$ or reflected to detector $D_2$ with amplitude $r$. Similarly for particles emitted from source $S_2$ (these processes are not illustrated in Figure). For photons, which obey Bose-Einstein statistics, the physical realisation of a beam splitter is a partially silvered mirror. In the case of electrons, which obey Fermi-Dirac statistics,  a beam splitter can be realized using edges of an integer  quantum Hall liquid impinging on a quantum point contact.}
\label{beam-splitter}
\end{figure}

In this Article we confine our attention to  Abelian anyons, emerging as 
elementary charged excitations of a FQH-state at a Laughlin filling $\nu$.  
We focus on a simple measurement with LQPs colliding at a beam splitter-like device, such an
experiment is not subject to some of the intricacies found in interferometric setups.
In order to illustrate our approach, 
we start by considering a simple, but instructive,  example. 
With reference to Fig.~\ref{beam-splitter}, we  consider a beam splitter 
where particles are injected from sources $S_1$ and $S_2$ and measured at 
detectors $D_1$ and $D_2$. An  incoming particle from  $S_1$  can be either 
transmitted to $D_2$ with scattering amplitude $t$, or reflected to $D_1$  
with scattering amplitude $r$ . Similarly an  incoming particle from  $S_2$  
can be either transmitted to $D_1$ with scattering amplitude $t'$, or 
reflected to $D_2$  with scattering amplitude $r'$, so that  
the scattering matrix describing these processes is given by 
\begin{equation}
S=\left(
\begin{array}{cc}
r & t' \\ 
t & r'
\end{array} \right)\,. 
\label{scamat}
\end{equation}
\noindent
Because of the particle number conservation, $S$ must be unitary, 
which enables us to use the following parametrization
\begin{equation}
S=\left(
\begin{array}{cc}
\sqrt{\mathcal{R}} & \sqrt{\mathcal{T}} \\  
\sqrt{\mathcal{T}} & -\sqrt{\mathcal{R}}
\end{array} \right) \, ,
\label{param}
\end{equation}
\noindent
with $\mathcal{T}$ and $\mathcal{R}$ respectively being the transmission and reflection 
coefficients. Let  $n_{D_1}$ and  $n_{D_2}$ respectively be the 
particle number operators  at $D_1$ and at $D_2$. Let us assume 
that the particles considered are either fermions, or bosons. 

Following Ref.[\onlinecite{Blanter:2000}], and considering a toy model with just one quantum mode per arm, one can calculate the correlation between the number of particles measured at $D_1$ and at $D_2$,  i.e.  $\langle \langle  n_{D_1}  n_{D_2} \rangle \rangle=\langle n_{D_1}  n_{D_2} \rangle -\langle n_{D_1}  \rangle \langle  n_{D_2} \rangle$. 
Notwithstanding that in optics very special incoming states can be realized,
in a typical  experiments, such as in transport measurements in solid-state physics---the appropriate tool
to investigate Abelian anyons at a FQH-edge---particles colliding at a  
beam splitter emerge from  thermal reservoirs. 

Assuming that the particles are emitted from two independent reservoirs $S_1$ and $S_2$, 
respectively  characterised by (thermal) distribution functions $n_1$ and $n_2$, one obtains  
\begin{equation}
\langle \langle  n_{D_1}  n_{D_2} \rangle \rangle=\pm \mathcal{RT} (n_1-n_2)^2,
\label{fermion-boson-correlation}
\end{equation}
\noindent
where the plus and minus sign refer to bosons and fermions respectively. 
It is worth stressing that, as it is apparent from Eq.(\ref{fermion-boson-correlation}), 
when $n_1 = n_2$,  the correlations  vanish, irrespectively of the underlying quantum 
statistics \cite{Buettiker:1991}. In this article we derive the analog 
of Eq.(\ref{fermion-boson-correlation}) for LQPs originating from sources (FQH edges) kept at 
the same temperature but, in general,  at different chemical potentials. For FQH anyons  there is no  
simple description of the beam splitter in terms of a scattering matrix. 
Therefore, we perform the calculation by resorting to  non-equilibrium Keldysh formalism. 
In particular, we realize the beam splitter as a quantum point contact (QPC) which allows
for LQP tunneling between the edges.   
Besides   weak inter-edge tunneling, no other 
approximation is involved in our calculation. As a result, while in 
the ''shot-noise'' regime, $| e V | \gg k_B T$ ($V$ being the voltage bias
between the edges and $k_B$ being the Boltzmann constant) we recover that  correlations 
are proportional to the tunneling current, 
with the constant of proportionality being equal to $q$, in the 
''thermal'' regime $| e V | \ll k_B T$ the  constant of proportionality
is renormalized by a purely statistics dependent function  $\gamma ( \nu ) = ( 6  / \pi^2) 
\partial_z^2 \ln \Gamma ( z ) \big|_{ z = \nu }$   [$\Gamma (z )$ being
Euler Gamma function], which can be  
directly measured by looking at current-current correlation probed in the appropriate 
regime. 
  
The Article is organized as follows:
In Section \ref{model} we introduce the model for a 
beam splitter realized with edge channels of a FQH system;
In Section \ref{curcorre},  we calculate the correlation of currents measured at different 
drains as  a function of the voltage bias V and 
the temperature $T$.
In Section \ref{statcur} we show how fractional statistics
can be probed from current-current correlation normalized to the tunneling 
current. 
In Section \ref{conclu} we  discuss and summarize our results and give an outlook 
of the possible implications of our work.
Mathematical details and a review of the non-interacting 
case ($\nu = 1$) are provided in the appendices.

\section{The Model}
\label{model}

In this Section we introduce the model for the edge channels that we use in the 
calculation of the current-current correlation. 
 
Throughout this Article, we limit our analysis to 
Laughlin's states at filling $\nu $, which are characterized by only 
one branch of chiral excitations per edge \cite{wenbook}.
This is not a potential limitations, 
as we outline in the concluding Section.  Our analysis, indeed, is expected to be
generalizable to non-Laughlin FQH states, e.g.  $\nu = 2 /3$ and $\nu=5/2$,
as a possible tool to investigate the properties of these more exotic FHQ states. 

\begin{figure}[h]
\includegraphics[width=0.9\linewidth]{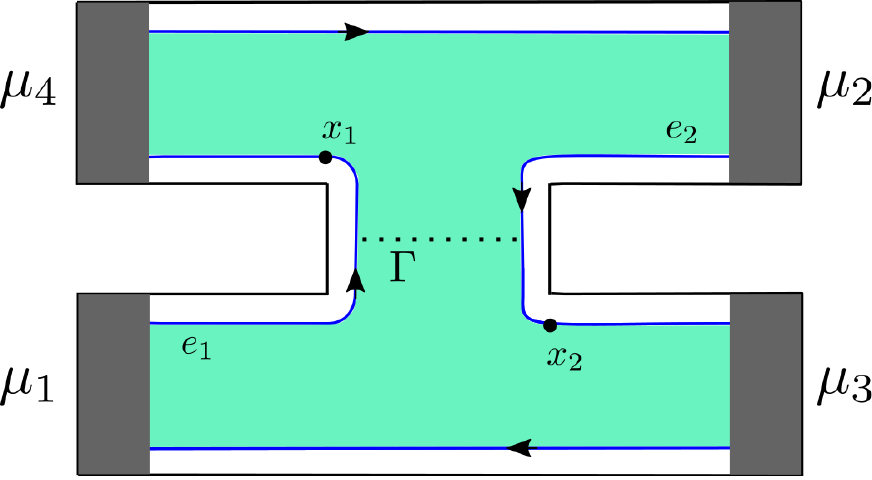}
\caption{In Figure we give a schematic representation of the device used for the proposed measurement. The green coloured area represents the incompressible electron liquid due to a strong perpendicular magnetic field. The boundary of the electron liquid (blue lines) are the edge channels supporting gapless excitations. As discussed in the main text we only need to focus on edges $e_1$ and $e_2$, they originate respectively from reservoirs at chemical potential $\mu_1$ and $\mu_2$. The dotted line represents tunneling between the two edges due to a quantum point contact. Currents are measured at points $x_1$ and $x_2.$}
\label{setup1}
\end{figure}

The device we discuss here has four edge channels (cfr. Fig.~\ref{setup1}), we 
only need to focus onto the ones labelled $e_1$ and $e_2$. In order to realize a beam splitter, 
we assume that a QPC  is obtained  between the two channels 
by means of electric gates, allowing for quasiparticle tunneling between $e_1$ and $e_2$.
Finally, it is worth stressing that we choose our  geometry  to allow for independent tuning of the 
chemical potentials at $e_1$ and $e_2$, respectively $\mu_1$ and $\mu_2$.

Edge excitations of Laughlin's FQH states are described within chiral Luttinger 
liquid (CLL)-framework \cite{wenbook}. In the two-edge model, the Hamiltonian for the 
edges is given by

 \begin{equation}
H_0=\frac{\hbar v}{4 \pi}\sum_{k=1,2} \int dx(\partial_x \phi_k(x))^2
\:\:\:\:, 
\label{fham}
\end{equation}
\noindent
with $v$ the plasmonic velocity. The chiral bosonic fields $\{  \phi_1(x) , \phi_2 (x)  \}$ obey the commutation
relations 
\begin{equation}
[\phi_k(x),\phi_l (x')]=i \pi \delta_{k , l} \mbox{sgn}(x-x')
\:\:\:\: .
\label{comrel}
\end{equation}
\noindent
With the normalizations in Eqs.(\ref{fham},\ref{comrel}), the   density 
operator at edge-$k$ $(k=1,2)$, $\rho_k ( x )$ is given by
 
\begin{equation}
\rho_k(x)=-\frac{\sqrt{\nu}}{2\pi}\partial_x \phi_k(x)
\, ,
\label{cdens}
\end{equation}
\noindent
while, because of the chiral propagation along the edges, the electric current density operator is:  $i_k ( x ) =e v \rho_k(x)$.  

The Hamiltonian operator describing  tunneling of a charge-$q$  LQP
at the QPC is constructed in terms of the quasiparticle creation and annihilation
operators at edge-$k$. Within CLL-framework, these are realized in terms of  vertex operators,
respectively given by 

\begin{eqnarray}
V_k^\dagger  ( x ) &=& F_k^\dagger e^{ i \sqrt{\nu } \phi_k ( x ) } \nonumber \\
V_k ( x ) &=& F_k e^{ - i \sqrt{\nu } \phi_k ( x ) } 
\:\:\: , 
\label{vertices}
\end{eqnarray}
\noindent
with $\{ F_k , F_k^\dagger \}$ being Klein factors that one has to introduce, in 
order to recover the correct commutation relations between operators belonging to 
different edges.
Choosing the  $x$-coordinates so that the QPC is located at $x=0$, we 
take the tunneling Hamiltonian to  be 
\begin{equation}
H_T=\Gamma V^\dag_1(0)V_2(0)+ {\rm h.c.}  
\end{equation}
We have assumed to work in a temperature/voltage regime such that  
terms that are less relevant  in the renormalization group sense \cite{Kane:1992,Kane:1994b} 
can be disregarded. These terms, indeed,
correspond to tunneling of quasiparticles with 
charge being an integer multiple of $q$.

In fact, as our device contains only one QPC, Klein factors can 
be dropped from the tunneling Hamiltonian $H_T$. Following Ref.[\onlinecite{Guyon:2002}],
the commutation rules between Klein factors must be assigned 
so that vertex operators corresponding to different edges must obey 
the same commutation relations as vertex operators corresponding to 
the same edge, that is,   $e^{i\sqrt{\nu}\phi_k(x_1)}  e^{i\sqrt{\nu}\phi_{k'} (x_2)} = 
e^{ i \pi \nu {\rm sgn} ( x_1 - x_2 ) } e^{i\sqrt{\nu}\phi_{k'} (x_2)} e^{i\sqrt{\nu}\phi_k(x_1)}$.
As a result, they have to satisfy the relations 
 $F_i^\dag F_i=F_i F_i^\dag=1$,  $F_1 F_2=e^{i\nu \pi} F_2 F_1$, and  
$F_1^\dag F_2=e^{-i\nu \pi} F_2 F_1^\dag $. Taking into account these commutation
relations, it is easy to check that the commutator between $H_T$ in interaction
representation computed at different times, that is, $ [ H_T ( t_1 ) , H_T ( t_2 ) ]$ is the 
same, whether or not one introduces the Klein factors in the vertex operators in Eq.(\ref{vertices}).
Therefore, they can safely disregarded, without affecting the validity of our derivation. 
The tunneling Hamiltonian $H_T$ can be simplified to 
\begin{equation}
H_T=\Gamma e^{i \sqrt{\nu} (\phi_1(0)-\phi_2(0) ) }+ {\rm h.c.}  
\:\:\:\: . 
\label{tunham}
\end{equation}
\noindent

The key quantity we consider in the following is the  
correlation function  between $i_1 (x_1 , t_1 )$ and $i_2 ( x_2 , t_2 )$, where 
$i_k ( x , t )$ is the current operator at edge $k$ in Heisenberg representation, 
$x_1 \in e_1$, $x_2 \in e_2$, and  both points $x_1$ and $x_2$ are situated  after  the 
QPC (in the sense of the propagation direction defined on each edge). Within 
the CLL-formalism, the correlation functions can be derived in a perturbative expansion 
in $\Gamma$, as we present in the next Section.

\section{Current-Current correlation}
\label{curcorre}

In this Section we illustrate the details of our calculations of the correlation of currents measured at the points $x_1$ and $x_2$ as function of the temperature and of the chemical potentials $\mu_1$ and $\mu_2$.

The finite frequency current-current correlation reads
\begin{multline}
S(\Omega;x_1,x_2)=\frac{1}{2}\int_{-\infty}^{+\infty}d(t_1-t_2)\langle \langle \hat{i}_1(t_1,x_1)\hat{i}_2(t_2,x_2)\\  +\hat{i}_2(t_2,x_2)\hat{i}_1(t_1,x_1)\rangle\rangle 
e^{i\Omega (t_1-t_2)} \,.
\label{correlation-anyons}
\end{multline}
Similar current-current correlation has been studied in the context of a quantum spin Hall system\cite{Lee:2012}.
Henceforth operators with a "hat"  are to be understood in the Heisenberg representation.
We first evaluate the finite frequency correlation $S(\Omega;x_1,x_2)$, later, taking the limit for  $\Omega\rightarrow 0$, correctly 
calculate $S(0)$ as it will be clear from the discussion below.

\begin{figure}[h]
\includegraphics[width=0.8\columnwidth]{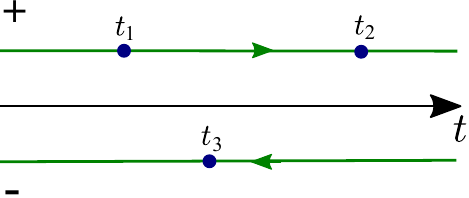}
\caption{Keldysh contour, the direction of the arrows indicates the ordering of times along
the contour. Here $+$ and $-$ indicate the upper and the lower branches, and will be used to define the  
four components of the Keldysh Green's function. As example of time ordering on the contour we have 
$t_2$ at later time than $t_1$ but at earlier time with respect to $t_3$.}
\label{keldyshcontour}
\end{figure}

Introducing the Keldysh time contour (see Fig.~\ref{keldyshcontour}), and the Keldysh time ordering operator $T_K$ we can rewrite the 
previous expression as 
\begin{multline}
S(\Omega;x_1,x_2)=\frac{1}{2}\sum_{\eta=\pm1}\int_{-\infty}^{+\infty}d(t_1-t_2)e^{i\Omega (t_1-t_2)}  \\ \times \langle \langle T_K i_1(t_1,x_1,\eta)i_2(t_2,x_2,-\eta)\rangle\rangle 
e^{i\Omega (t_1-t_2)} \,.
\end{multline}
Notice that in the above equation we have introduced an index $\eta=\pm 1$ which specify the upper and the lower part of the Keldysh contour.

We assume that the tunneling $H_T$ is adiabatically turned on at $t=-\infty$. In order to evaluate Eq.(\ref{correlation-anyons}) we move to the interaction representation with respect to $H_0$, and rewrite Eq.(\ref{correlation-anyons}) as
\begin{multline}
S(\Omega;x_1,x_2)=\frac{1}{2}\sum_{\eta=\pm1}\int_{-\infty}^{+\infty}d(t_1-t_2) e^{i\Omega (t_1-t_2)} \\
\langle \langle T_K i_1(t_1,x_1,\eta)i_2(t_2,x_2,-\eta)S_K\rangle\rangle  \,,
\end{multline}
where $S_K=T_K\exp\{-\frac{i}{\hbar}\int_K H_T(\tau)d\tau\}$ with $K$ labelling  the Keldysh contour.
Notice that operators without the "hat" are to be understood in the interaction representation with respect to $H_0$.
Expanding $S_K$ to the lowest non vanishing order in the tunnelling Hamiltonian we have 
\begin{multline}
S(\Omega;x_1,x_2)= 
-\frac{1}{4 \hbar^2}\sum_{\eta,\eta_1,\eta_2=\pm1}\eta_1 \eta_2   \\ \times    \int_{-\infty}^{+\infty}ds_1 \int_{-\infty}^{+\infty}ds_2 
 \int_{-\infty}^{+\infty}d (t_1-t_2)    e^{i\Omega (t_1-t_2)}    \\ \times  
\langle \langle T_K i_1(t_1,x_1,\eta)i_2(t_2,x_2,-\eta)H_T(s_1,\eta_1) H_T(s_2,\eta_2)\rangle\rangle  \,.
\end{multline}
Keeping only connected contributions and dropping terms that are trivially zero by Keldysh integration 
we may rewrite the previous expression as:

\begin{multline}
S(\Omega;x_1,x_2)=-\frac{|\Gamma|^2e^2 \nu v^2}{16\pi^2 \hbar^2}
\sum_{\eta,\eta_1,\eta_2=\pm1}\eta_1 \eta_2  \times \\
  \int_{-\infty}^{+\infty}ds_1 \int_{-\infty}^{+\infty}ds_2  \int_{-\infty}^{+\infty}d (t_1-t_2)    e^{i\Omega (t_1-t_2)} 
\Big\{
\\ 
\langle T_K \partial_x\phi_1(x_1,t_1,\eta)e^{i\sqrt{\nu}\phi_1(0,s_1,\eta_1)}e^{-i\sqrt{\nu}\phi_1(0,s_2,\eta_2)}  \rangle
\\ \times  \langle T_K \partial_x\phi_2(x_2,t_2,-\eta)e^{-i\sqrt{\nu}\phi_2(0,s_1,\eta_1)}e^{i\sqrt{\nu}\phi_2(0,s_2,\eta_2)}  \rangle
\\ + \, 
\langle T_K \partial_x\phi_1(x_1,t_1,\eta)e^{-i\sqrt{\nu}\phi_1(0,s_1,\eta_1)}e^{i\sqrt{\nu}\phi_1(0,s_2,\eta_2)}  \rangle
\\ \times  \langle T_K \partial_x\phi_2(x_2,t_2,-\eta)e^{i\sqrt{\nu}\phi_2(0,s_1,\eta_1)}e^{-i\sqrt{\nu}\phi_2(0,s_2,\eta_2)}  \rangle \Big\} \,.
\label{curr-curr1}
\end{multline}
In order to explicitly compute the  multiple correlators at finite 
$\mu_1$ and $\mu_2$ entering Eq.(\ref{curr-curr1}), we recall that, adding a nonzero chemical potential
$\mu$ to a chiral Luttinger liquid described by the Hamiltonian of Eq.(\ref{fham}) is equivalent (apart for an over-all constant contribution 
to the groundstate energy) to the replacement $\partial_x \phi (x ) \to  
\partial_x \phi (x ) - \frac{\mu \sqrt{\nu}}{v}$. Therefore, denoting with 
$\langle \ldots \rangle_0$ the averages computed at $\mu_1 = \mu_2 = 0$, we obtain 
\begin{multline}
\label{finite-potential}
\langle
T_K \partial_x\phi_k(x,t,\eta)e^{i \sqrt{\nu}\phi_k(0,t_k,\eta_1)}
e^{-i \sqrt{\nu}\phi_k(0,t_2,\eta_2)}
\rangle_{\mu_k}= \\
\Big\{
\langle
T_K \partial_x\phi_k(x,t,\eta)e^{i \sqrt{\nu}\phi_k(0,t_1,\eta_1)}
e^{-i \sqrt{\nu}\phi_k(0,t_2,\eta_2)}
\rangle_0 +  \\ 
-\frac{\mu_k \sqrt{\nu}}{v}\langle T_Ke^{i \sqrt{\nu}\phi_k(0,t_1,\eta_1)}
e^{-i \sqrt{\nu}\phi_k(0,t_2,\eta_2)} \rangle_0 \Big\}e^{i \nu \mu_k(t_1-t_2)} \,,
\end{multline}
and similarly for the conjugate expression. Notice that contributions to Eq.(\ref{curr-curr1}) proportional to the chemical potentials $\{\mu_k\}$ vanish identically 
after integration over the Keldysh contour.
In order to complete the calculation we can use the following identity
\begin{multline}
\langle
T_K \partial_x\phi_k(x,t,\eta)e^{i \sqrt{\nu}\phi_k(0,t_1,\eta_1)}
e^{-i \sqrt{\nu}\phi_k(0,t_2,\eta_2)}
\rangle= \\ 
 -i \, \partial_x \lim_{\lambda \rightarrow 0} \partial_\lambda \langle
T_K e^{i \lambda \phi_k(x,t,\eta)}e^{i \sqrt{\nu}\phi_k(0,t_1,\eta_1)}
e^{-i (\sqrt{\nu}+\lambda)\phi_k(0,t_2,\eta_2)}
\rangle \,.
\end{multline}
We finally obtain 
\begin{eqnarray}
&& \langle
T_K \partial_x\phi_{k } (x,t,\eta)e^{i \sqrt{\nu}\phi_{k  } (0,t_1,\eta_1)}
e^{-i \sqrt{\nu}\phi_{k  } (0,t_2,\eta_2)}
\rangle_0 = \nonumber   \\
&& \frac{\sqrt{\nu}\pi}{\hbar \beta v} \Big(\cot\left\{ \frac{\pi}{\hbar \beta}
\left[ i(t-t_1-x/v)+\tau_c \, \sigma_{\eta,\eta_1}(t-t_1)\right]\right\} \nonumber \\
&& -\cot\left\{ \frac{\pi}{\hbar \beta}\left[ i(t-t_2-x/v)+\tau_c \, 
\sigma_{\eta,\eta_2}(t-t_2)\right]\right\} \Big) \nonumber 
\\ && \times G^{(\nu)}_{\eta_1,\eta_2}( t_1 - t_2 ) 
\label{density-insertion}
\:\:\:\: . 
\end{eqnarray}
\noindent
In Eq.(\ref{density-insertion}), we have set  $\beta = ( k_B T )^{-1}$.  Also, we have defined 
$\sigma_{\eta,\eta'}(t-t')=[(\eta+\eta'){\rm sgn} (t-t')+\eta'-\eta]/2$ and 
have introduced the   cutoff
time $\tau_c = l_c / v$, with $l_c$ being a short-distance cutoff length.
Moreover, we have introduced the Keldysh Green function  
$G^{(\nu)}_{\eta_1,\eta_2}( t_1 - t_2 ) =
 \langle  T_K   e^{ i \sqrt{\nu} \phi_{k  }  (0, t_1 , \eta_1 ) } 
 e^{ - i \sqrt{\nu}  \phi_{ k  }  (0, t_2 , \eta_2 ) }  \rangle_0$, 
given by
 \begin{eqnarray}
&& G^{(\nu)}_{\eta_1,\eta_2}( t_1 - t_2 ) =l_c^\nu \nonumber \\
&& \times \left( \frac{\hbar \beta v}{\pi}\sin\left\{ \frac{\pi}{\hbar \beta}
[i(t_1-t_2)\sigma_{\eta_1,\eta_2}(t_1-t_2)+\tau_c]\right\}\right)^{-\nu} 
\label{greenfunc.1}
 \end{eqnarray}
\noindent

The cutoff-dependent contribution to the argument of the cotangent functions 
at the second  and at the third line of Eq.(\ref{density-insertion}) is 
effective only when $t-t_1-x/v \sim 0$ (second line), or when 
$t-t_1-x/v \sim 0$ (third line). This enables us to set  $\sigma_{\eta,\eta_1}(t-t_1)
= \sigma_{\eta,\eta_1}( x / v ) = \eta_1$ (second line), and 
$\sigma_{\eta,\eta_2}(t-t_2) = \sigma_{\eta,\eta_2}( x / v ) = \eta_2$ (third line). 
As a result, we may eventually rewrite Eq.(\ref{density-insertion}) as 

\begin{eqnarray}
&& \langle
T_K \partial_x\phi_{k } (x,t,\eta)e^{i \sqrt{\nu}\phi_{k  } (0,t_1,\eta_1)}
e^{-i \sqrt{\nu}\phi_{k  } (0,t_2,\eta_2)}
\rangle_0 = \nonumber   \\
&& \frac{\sqrt{\nu} }{  v} \left[ \xi_{ \eta_1} \left( t - t_1  - \frac{x}{v} 
\right) -  \xi_{ \eta_2} \left( t - t_2  - \frac{x}{v} 
\right)  \right]   \nonumber  
\\ && \times G^{(\nu)}_{\eta_1,\eta_2}( t_1 - t_2 ) 
\label{density-insertion2}
\:\:\:\: , 
\end{eqnarray}
\noindent
with 
\begin{equation}
\xi_\eta ( t ) = \frac{\pi}{\hbar\beta} \cot \left[ \frac{\pi}{\hbar\beta} 
( i t + \eta \tau_c ) \right] \,\:\:\:.
\label{phieta}
\end{equation}
\noindent
 
\noindent
 Taking into account the result in Eq.(\ref{density-insertion2}), it is now
 
possible to explicitly compute $S( \Omega ; x_1 , x_2 )$.  Introducing the
Fourier transform of $ G_{ \eta_1 , \eta_2}^{( \nu)}$ and of $\xi_\eta$ (see Appendix \ref{grenfu} for details)
we obtain
\begin{multline}
 S( \Omega ; x_1 , x_2 ) = \frac{ | \Gamma |^2 e^2 \nu^2}{4 \pi^2\hbar^2} \: 
 e^{ i \Omega ( x_1 - x_2 ) } 
  \sum_{ \eta_1 , \eta_2 } \eta_1 \eta_2    \\ \times \biggl\{ [ \xi_{\eta_1} ( \Omega )   
 \xi_{\eta_1} ( - \Omega ) +
 \xi_{\eta_2} ( \Omega )   \xi_{\eta_2} ( - \Omega ) ] 
 G_{ \eta_1 , \eta_2}^{(2 \nu)}  ( \nu \Delta \mu )   \\ -
  \xi_{\eta_1} ( \Omega ) \xi_{ \eta_2} ( - \Omega ) 
  G_{ \eta_1 , \eta_2}^{(2 \nu)}  ( \nu \Delta \mu + \Omega ) 
 \\ - 
 \xi_{\eta_1} ( - \Omega ) \xi_{ \eta_2} (   \Omega ) 
 G_{ \eta_1 , \eta_2}^{(2 \nu)}  ( \nu \Delta \mu   -\Omega ) 
 \biggr\} \, ,
 \label{curr-curr2}
\end{multline}
\noindent
with $\Delta \mu = \mu_1 - \mu_2$. Using the 
explicit formulas for  $G_{\eta_1 \eta_2}^{(\nu)}$ and $\xi_\eta ( \Omega )$ (see Eqs.(\ref{ftra.11},\ref{ftra.12},\ref{fad.1},\ref{fad.2},\ref{cor.a2}) ) we perform the sum over the Keldysh indices. Taking the limit for $\Omega \rightarrow 0$ eventually we obtain
\begin{multline} \label{curr-curr3}
S(0)=2 i\left( \frac{ \hbar \beta }{2 \pi} \right)^{1 - 2 \nu} 
 \frac{ e^2 \nu^2 \tau_c^{2\nu}| \Gamma |^2}{\hbar^2\pi \Gamma ( 2 \nu) } 
 \: \sinh \left( \frac{\beta \nu \Delta \mu}{2} \right)
\\
 \times
\left| \Gamma \left( \nu + \frac{i \beta \nu \Delta \mu}{2 \pi} \right)\right|^2
 \\
 \times \left[ \psi(\nu + \frac{i \beta \nu \Delta \mu}{2 \pi} ) 
 -\psi(\nu - \frac{i \beta \nu \Delta \mu}{2 \pi} )\right] \,.
 \end{multline}
 \noindent
To recover a compact notation, in Eq.(\ref{curr-curr3}) we  have expressed 
$S ( 0 )$ in terms of Euler Gamma function $\Gamma ( z )$ and of 
its logarithmic derivative, the digamma function $\psi ( z ) = \partial_z [ \ln \Gamma ( z ) ]$.  
As anticipated, in order to obtain the correct  result for $S(0)$, one has
to first perform the calculation of $ S ( \Omega; x_1 , x_2  )$ at finite $\Omega$ and 
then take the limit $\Omega \to 0$ at the end of the calculation, thus avoiding 
problems related to $\xi_\eta ( \Omega )$ being ill-defined as $\Omega \to 0$ (see appendix
\ref{grenfu} for details). Taking  $\nu=1$ in Eq.(\ref{curr-curr3}) reproduces the known 
result for non-interacting electrons\cite{Buettiker:1992} which we
discuss in detail in Appendix \ref{chifer}. 

In the next Section we will look at the ratio $S(0)/i_T$, with $i_T$ being the tunneling current across the QPC. For the reader's 
convenience we report below the standard result\cite{Wen:1991} 
\begin{multline}
i_T=
\frac{2 q}{\hbar^2\Gamma(2\nu)}|\Gamma|^2 \tau_c^{2\nu}
\left(\frac{\hbar\beta}{2\pi}\right)^{1-2\nu}
\\ \times \sinh\left( \frac{\beta \nu \Delta\mu}{2}\right) 
\left| \Gamma\left( \nu+i\frac{\beta\Delta\mu \nu}{2\pi}\right)\right|^2 
\, .
\label{tuncor}
\end{multline}
The  generalization of Eq.(\ref{tuncor}), beyond perturbative expansion,  was calculated exactly by Bethe-ansatz in Refs.[\onlinecite{Fendley:1995,Fendley:1995c,Fendley:1996}].

\section{Fractional statistics detection from current-current correlation}
\label{statcur}
Equation (\ref{curr-curr3}) is  the  main result of this Article, in this Section we discuss
its consequences.

As a first comment we notice that in analogy with 
the result in Eq.(\ref{fermion-boson-correlation}),  we find that $ S ( 0 ) = 0$ 
for $\mu_1=\mu_2$. This is consistent with B\"uttiker's  result 
of Ref.[\onlinecite{Buettiker:1992}] for the non-interacting case ($\nu=1$), 
and it is now generalized to the case of Laughlin fractions. Such a result 
for $\nu=1$ is easily shown to be in agreement with the 
fluctuation-dissipation theorem. Indeed for non-interacting electrons
the equilibrium correlation function between 
currents  measured at drains $\alpha$ and $\beta$, $i_\alpha$ and $i_\beta$, 
satisfies the relation 
\begin{equation}
  \int_{ - \infty}^{ + \infty} \:  \langle\langle \{i_\alpha(t), i_\beta(0) \}
\rangle \rangle dt =2({\bf G}_{\alpha , \beta}+{\bf G}_{\beta , \alpha}) k_B T
\;\; , 
\label{fdtheo}
\end{equation}
\noindent
with  $\bf{G}_{\alpha ,\beta}$ being the dc conductance   
between terminals $\alpha$ and  $\beta$. In fact, in the particular geometry 
we are considering here, no electric current can flow
between   $x_1$ and $x_2$,   because of the chiral propagation along 
the edge channels. Therefore,  we have shown that a result similar to that of Ref.[\onlinecite{Buettiker:1992}] 
also applies to currents at the edges of  a FQH liquid. Moreover,  Eq.(\ref{fermion-boson-correlation}) 
shows that, for any Laughlin filling $\nu$, the current-current correlation is negative, 
suggesting that the beam splitter geometry we consider highlights the exclusion statistics character 
of Laughlin's quasi-particles\cite{Haldane:1991}. This result agrees with Ref.[\onlinecite{Safi:2001}] for  their case $\nu=1/3$ but not for 
$\nu\le1/5$, and it is in contrast with Refs.[\onlinecite{Kim:2005,Campagnano:2012,Campagnano:2013}]. 
We suspect  that  this is somehow related to the different geometries involved, and we will investigate this issue in future works. 

We also notice that negative correlations are found in Ref.[\onlinecite{Rosenow:2015}]  where current-current correlation is studied for a beam of diluted anyons impinging on a beam splitter, analysis complementary to the study reported here. 

Besides the results outlined above, our most important finding is 
that, combining together   Eq.(\ref{curr-curr3}) and  
Eq.(\ref{tuncor}), it is possible to propose a way to directly measure the fractional
statistics of LQPs.  A key observation is now that, by setting $\Delta \mu = e  V$, where
$V$ is the voltage bias between $e_1$ and $e_2$, the argument of the Gamma and the digamma functions 
in Eqs.(\ref{curr-curr3},\ref{tuncor}) can be rewritten as $\zeta = \nu + i \frac{\beta q V}{2 \pi}$. 
Roughly speaking, one might say that $\Re e ( \zeta )$ carries information about the 
fractional statistics, while $\Im m ( \zeta )$ carries information about the 
fractional charge. Therefore, one might expect that either information can 
be extracted, according to whether one considers the formulas in the 
limit $ | \Re e ( \zeta ) / \Im m ( \zeta ) | \gg 1$, or 
$ | \Re e ( \zeta ) / \Im m ( \zeta ) | \ll 1$. 

Let us discuss first the case $ | \Re e ( \zeta ) / \Im m ( \zeta ) | \ll 1$,
corresponding to $| e V | \gg k_B T$. In this regime, an appropriate approximation 
for Eqs.(\ref{curr-curr3},\ref{tuncor}) can 
be derived by using Stirling's formula for the $\Gamma$-functions, that is

\begin{equation}
 \Gamma ( z ) \approx \sqrt{2 \pi} \: ( z -1 )^{ z - \frac{1}{2} } \: e^{ - z + 1} 
 \:\:\:\: ,
 \label{stirling}
\end{equation}
\noindent
valid for $ | z | \gg 1$. Using Eq.(\ref{stirling}), one finds the following 
asymptotic expansions for $ S ( 0 )$ and $i_T$ (assuming $V > 0$)
\begin{eqnarray}
  S ( 0 ) & \approx & - \frac{2 \pi q^2 | \Gamma |^2 \tau_c^{ 2 \nu} }{\hbar^{2\nu+1}~ \Gamma [ 2 \nu ] }
   ( q V )^{ 2 \nu - 1 } \nonumber \\
    i_T & \approx &  \frac{2 \pi q | \Gamma |^2 \tau_c^{ 2 \nu} }{\hbar^{2\nu+1}
   \Gamma [ 2 \nu ] }
  ( q V )^{ 2 \nu - 1 }
   \:\:\:\: .
   \label{f.9}
\end{eqnarray}
\noindent
Eqs.(\ref{f.9}) suggest that, for  $| e V | \gg k_B T$, the fractional charge $q$
can be directly probed  by looking at the ratio $ q = | S ( 0 ) | / i_T $ between
two directly measurable quantities such as $S ( 0 )$ and $i_T$, which is 
the main idea typically implemented in shot-noise based measurements of the 
fractional charge (notice that here, instead, we look at correlation 
between currents at different drains). 

In the complementary limit, in order to directly access to informations on the fractional statistics, one 
has rather to consider the thermal regime, namely, 
$| e V | \ll k_B T$. In this regime, the limiting formulas for  Eqs.(\ref{curr-curr3},\ref{tuncor}) 
can be recovered by expanding the Gamma and the digamma functions to leading order in 
$\Im m ( \zeta ) / \Re e ( \zeta )$, obtaining 
\begin{small}
\begin{eqnarray}
 S ( 0 ) &\approx& - \frac{q^2 | \Gamma |^2 \tau_c^{ 2 \nu}  }{ \hbar^{2\nu+1}~\Gamma [ 2 \nu ] }
 \left( \frac{\beta}{2 \pi} \right)^{ 1 - 2 \nu }   ( \beta q V )^2
 \{ \Gamma^{''} [ \nu ] \Gamma [ \nu ] - ( \Gamma^{'} [ \nu ])^2 \} \, ,
\nonumber \\ 
  i_T &\approx&  \frac{q | \Gamma |^2 \tau_c^{ 2 \nu} \Gamma^2 [ \nu ] }{\hbar^{2\nu+1}~ \Gamma [ 2 \nu ] }
 \left( \frac{\beta}{2 \pi} \right)^{ 1 - 2 \nu }  ( \beta q V  ) 
  \, .
 \label{f.10}
\end{eqnarray}
\end{small}
\noindent
From Eqs.(\ref{f.10}) one therefore obtains 

\begin{equation}
 \frac{ | S ( 0 ) |}{i_T } = \left[ \frac{q^2 \pi^2V}{6 k_B T} \right] 
 \gamma ( \nu ) 
\:\:\:\:, 
\label{f.11}
\end{equation}
\noindent
with $\gamma ( \nu ) = ( 6  / \pi^2)  \partial_z^2 \ln \Gamma ( z ) \big|_{ z = \nu }   $. 
Except for the factor $\gamma ( \nu )$, the result in Eq.(\ref{f.11}) is the same
one would obtain for non-interacting electrons ($\nu = 1$) by simply replacing
$e$ with $q$. Therefore, the  additional factor $\gamma ( \nu )$ is not a feature simply 
related to the fractional charge of LQPs---it is a clear signature of the 
quasi-particle {\em fractional statistics} which, as we propose, can be 
directly measured by looking at currents correlations probed in the appropriate 
thermal regime. We report here, for the readers' convenience, some numerical values of $\gamma(\nu)$, $\gamma(1)=$1, $\gamma(1/3)\simeq 6.18$, $\gamma(1/5)\simeq15.97$. 
 
 As a final remark, we notice that the reason to look at the correlation of currents measured at different drains
lies in the fact that, if one considers noise of the tunneling current, i.e. 
$S_{i_T}=(1/2)\int dt \langle \langle \{ i_T(0),i_T(t)\}  \rangle \rangle$, one would obtain\cite{Kane:1994b}
\begin{multline}
S_{i_T}=
\frac{2 q^2}{\hbar^2\Gamma(2\nu)}|\Gamma|^2 \tau_c^{2\nu}
\left(\frac{\hbar\beta}{2\pi}\right)^{1-2\nu}
\\ \times \cosh\left( \frac{\beta \nu \Delta\mu}{2}\right) 
\left| \Gamma\left( \nu+i\frac{\beta\Delta\mu \nu}{2\pi}\right)\right|^2 
\, .
\label{tun-cor-noise}
\end{multline}
Such a quantity normalized  to the tunneling current $i_T$ of Eq.(\ref{tuncor}), i.e. the Fano factor, only carries information about the quasi-particles' charge but not their statistics. 
\section{Summary and Outlook}
\label{conclu}
We have discussed the correlation of currents measured at separate drains in a beam splitter-like
geometry for fractional quantum Hall systems at Laughlin filling factors. 
Because of the chiral propagation of LQPs along the edge channels we have proved (within perturbation theory)
that the equilibrium correlation, i.e. for $\mu_1=\mu_2$, is zero, as it was found for chiral fermions ($\nu=1$). Using 
Keldysh technique we have also obtained expressions for the stationary out of equilibrium case and show how a correlation
measurements carries information about the fractional statistics. Our findings suggest an anti-bunching character of the  LQPs.  

In perspective our result might also provide a useful tool to investigate more exotic filling fractions like for instance 
$\nu=2/3$ where neutral counter-propagating modes have been predicted\cite{Kane:1994,Kane:1995} and recently 
observed\cite{Bid:2010,Yaron:2012,Gurman:2012}, but still in the need of a thorough characterisation. 

In such systems even  for both the measuring points $x_1$ and $x_2$ situated after the QPC, due to the counter-propagating modes and their interaction with the charge modes, one might 
expect a signal propagation between these two points---giving rise to a non zero {\em equilibrium} correlation. 
We will to investigate this possibility as tool to study counter-propagating modes in future works.
In addition, we also plan to expand our work to analyze the relation between correlations and fractional statistics-related interaction among particles with fractionalized quantum numbers\cite{Giuliano:2001,Giuliano:2001b,Giuliano:2003}.

\section*{Acknowledgements}
One of the authors (GC) kindly acknowledges inspiring discussions with Y. Gefen and I.V. Gornyi. We also thank M.  Koch-Janusz, A. Stern, A. Tagliacozzo, and Yi Zhang for useful  exchange of views. Financial support from MIUR-FIRB 2012 project HybridNanoDev (Grant No.RBFR1236VV) is gratefully acknowledged.
\appendix 

\section{ Green's functions}
\label{grenfu}
In this Appendix, we provide the details of the calculation of the
Green's functions we use  to compute the current-current
correlation, and their corresponding Fourier transforms. 
The first quantities we need are the correlation functions
of vertex operators evaluated on branches $\eta_1 , \eta_2$ of the Keldysh
path, 
\beq
G_{\eta_1 , \eta_2}^{(\nu) } ( t_1 , t_2 ) =
 \langle  T_K   e^{ i \sqrt{\nu} \phi_{ k  }  ( t_1 , \eta_1 ) } 
 e^{ - i \sqrt{\nu}  \phi_{ k }  ( t_2 , \eta_2 ) } \rangle_0
 \;\;\;\; .
\label{nov.1}
\eneq
\noindent  
A standard bosonization calculation yields the result in Eq.(\ref{greenfunc.1}), 
\begin{equation}
G_{\eta_1 , \eta_2}^{(\nu) } ( t_1,t_2 ) =  l_c^\nu \left( \frac{\hbar\beta v}{\pi} \right)^{ -  \nu} 
 \sin^{ -\nu} \left[ \frac{\pi}{\hbar \beta } ( i t \,\sigma_{\eta_1 , \eta_2} ( t_1-t_2 ) 
+ \tau_c ) \right] \:\:\:\: . 
\end{equation}
\noindent
For the sake of clarity, we list the Keldysh Green functions 
corresponding to the four possible choice of the Keldysh indices:
\begin{eqnarray}
 G_{+,+}^{(\nu) } ( t ) &=& l_c^\nu \left( \frac{\hbar \beta v}{\pi} \right)^{ -  \nu} 
\: \sin^{-\nu} \left[ \frac{\pi}{\hbar \beta} ( i | t |  + \tau_c ) \right] \nonumber ,\\
G_{-,-}^{(\nu) }  ( t ) &=& l_c^\nu \left( \frac{\hbar \beta v}{\pi} \right)^{ -   \nu} 
\: \sin^{ -   \nu} \left[ \frac{\pi}{\hbar \beta} ( - i | t |  + \tau_c ) \right] \nonumber, \\
 G_{-,+}^{(\nu) } ( t ) &=&  l_c^\nu\left( \frac{\hbar \beta v}{\pi} \right)^{ -   \nu} 
\: \sin^{ -   \nu} \left[ \frac{\pi}{\hbar \beta} ( i   t    + \tau_c ) \right] \nonumber ,\\
 G_{+,-}^{(\nu) } ( t ) &=& l_c^\nu \left( \frac{\hbar \beta v}{\pi} \right)^{ -  \nu} 
\: \sin^{ -   \nu} \left[ \frac{\pi}{\hbar \beta} ( - i   t    + \tau_c ) \right]
\,.
\label{cor.5}
\end{eqnarray}
\noindent
Next we compute the Fourier transform of Eqs.(\ref{cor.5}) defined as, 
\[ 
G_{ \eta_1 , \eta_2}^{(\nu) } ( \omega ) = 
\int_{-\infty}^{+\infty}  dt \,  e^{ i \omega t} \,  G_{ \eta_1 , \eta_2}^ {(\nu) }  ( t )
\:\:\:\: . 
\]
\noindent
In computing $G_{ \eta_1 , \eta_2}^{(\nu) } ( \omega )$, it is useful to start with 
$G_{ - , +}^{(\nu) } ( \omega )$ and with  $G_{ + , -}^{(\nu) } ( \omega )$. Moreover, 
in view of the identity $G_{ - , +}^{(\nu) } ( \omega )  = G_{ + , -}^{(\nu) } ( - \omega )$
(which is readily proved  from the definition of the Keldysh Green functions), 
one concludes that it is enough to just compute $G_{ - , +}^{(\nu) } ( \omega ) $.
In order to do so, we   notice that 
the branch points of $G_{ - , +}^{(\nu) } ( t) $ are located at 
$t_n = i ( \tau_c + \hbar \beta n )$, with $n = 0 , \pm 1 , \ldots $. Therefore, 
to make sure that 
no branch cuts intersect the real axis in computing $G_{ - , +}^{(\nu) } ( \omega ) $, 
we chose the phase branch so that $- \pi \leq {\rm arg} ( it ) < \pi$ and, accordingly, the branch cuts are all 
horizontal. Having stated this, $G_{ - , +}^{(\nu) } ( \omega )$ 
takes the following integral representation
\begin{multline}
 G_{- , +}^{(\nu) }  ( \omega ) =\\ l_c^\nu \left( \frac{\hbar \beta v}{2 \pi} \right)^{ - \nu } 
 \int_{ - \infty}^\infty \: d t \: e^{ i \omega t} 
  [ (-i) e^{ - \frac{\pi  t}{\hbar \beta } }e^{ i  \delta  } 
 - (-i) e^{ \frac{\pi t}{\hbar \beta} } e^{ - i  \delta } ]^{- \nu} 
\\ =   \frac{2 l_c^\nu}{v^\nu} \left( \frac{\hbar \beta }{2 \pi } \right)^{1 - \nu} 
 \int_0^\infty \: d u \: e^{ i \frac{\hbar \omega \beta}{\pi} u } [ 
 (-i) e^{ - u } e^{ i  \delta  } - (-i) e^u e^{ - i  \delta } ]^{- \nu} 
   \\
 +  \frac{2l_c^\nu}{v^\nu} \left( \frac{\hbar \beta }{2 \pi } \right)^{1 - \nu} 
 \int^0_{- \infty }\: d u \: e^{ i \frac{\hbar \omega \beta}{\pi} u } [ 
 (-i) e^{ - u } e^{ i  \delta  } - (-i) e^u e^{ - i  \delta}  ]^{-  \nu} 
  \\
 =   \frac{2 l_c^\nu}{v^\nu} \left( \frac{\hbar \beta }{2 \pi } \right)^{1 - \nu}  
 e^{ i  \pi \nu /2  }
 \int_0^\infty \: d u \: e^{ \left( i \frac{\hbar \omega \beta}{\pi} -   \nu  \right) u } 
 [ 1 - e^{ - 2 u } ]^{-  \nu}  \\
 +  \frac{2 l_c^\nu}{v^\nu} \left( \frac{\hbar \beta }{2 \pi } \right)^{1 -  \nu}   e^{ - i  \pi \nu / 2 }
 \int^0_{-\infty } \: d u \: e^{ \left( i \frac{\hbar \omega \beta}{\pi}+    \nu \right) u } 
 [ 1 - e^{   2 u } ]^{-  \nu}  \,. 
\label{ftra.6}
\end{multline}
\noindent
In Eq.(\ref{ftra.6}) we have set 
  $\delta =  \pi \tau_c / \hbar \beta$ and have taken advantage of 
the fact that,  in the last two lines, it was possible 
to drop the terms depending on  the regularizator $\delta$. 
Going through straightforward manipulation we can readily trade
Eq.(\ref{ftra.6}) for a known integral representation 
of the Beta-function, that is

\beq
B ( x , y ) = \frac{\Gamma ( x ) \Gamma ( y ) }{\Gamma ( x + y ) } = 
\int_0^1 \: d w \: w^{x-1} ( 1 - w )^{y-1}
\, .
\label{ftra.8}
\eneq
\noindent
To do so, we resort to the   integration  variable $w = e^{ - 2 u}$  ($w=e^{ 2 u }$ )  in the first (second) 
integral of Eq.(\ref{ftra.6}), so that we eventually obtain 

\begin{multline}
 G_{- , +}^{(\nu) }  ( \omega ) =\\ \left(\frac{l_c}{v}\right)^{\nu}
 \left( \frac{\hbar \beta }{2 \pi } \right)^{1- \nu} e^{ - i \pi   \nu / 2}
\: \int_0^1 \: d w \: w^{ i \frac{\hbar \omega \beta}{2 \pi} + \frac{\nu}{2}  - 1} ( 1 - w )^{- 
 \nu }  \\
+ \left(\frac{l_c}{v}\right)^{\nu}
 \left( \frac{\hbar \beta }{2 \pi } \right)^{1-  \nu}  e^{   i \pi\nu / 2}
\: \int_0^1 \: d w \: w^{ - i \frac{\hbar \omega \beta}{2 \pi} + 
\frac{\nu}{2}  - 1} ( 1 - w )^{-  \nu} 
\, . 
\label{ftra.7}
\end{multline}
\noindent
Comparing Eq.(\ref{ftra.7}) to Eq.(\ref{ftra.8}), we eventually
find 
\begin{multline}
 G_{- , +}^{(\nu) } ( \omega ) =  \\ \left(\frac{l_c}{v}\right)^{\nu}\left( \frac{\hbar\beta }{2 \pi } \right)^{1- \nu} 
 \Gamma ( 1 -  \nu ) \left[ e^{ i \pi \nu  / 2} 
 \frac{\Gamma \left( \frac{\nu}{2}  - i \frac{\hbar \omega \beta}{2 \pi}
 \right)}{\Gamma \left( 1 - \frac{ \nu }{2}  - i 
 \frac{\hbar \omega \beta}{2 \pi} \right) } \right.+\\ \left.  
 e^{ - i \pi \nu / 2} \frac{\Gamma \left( \frac{ \nu }{2}  + 
 i \frac{\hbar \omega \beta}{2 \pi}
 \right)}{\Gamma \left( 1 - \frac{ \nu }{2}  + 
 i \frac{\hbar \omega \beta}{2 \pi} \right) } \right]
 \, . 
 \label{ftra.9}
 \end{multline}
 \noindent
 Finally, using the identity 
  \beq
 \Gamma ( z ) \Gamma ( 1 - z ) = \frac{\pi}{\sin ( \pi z ) }
 \:\:\:\: , 
 \label{ftra.10}
 \eneq
 \noindent
 we can recast Eq.(\ref{ftra.9}) into the form 
 \beq
 G_{- , +}^{(\nu) } ( \omega ) = \frac{l_c^\nu}{v^\nu\Gamma (  \nu) }
 \left( \frac{\hbar\beta }{2 \pi  } \right)^{1 - \nu } 
 e^{  \frac{\hbar \omega \beta}{2} }  \left| \Gamma 
 \left( \frac{\nu }{2}  +  i \frac{\hbar \omega \beta}{2 \pi} \right)  \right|^2
 \, . 
  \label{ftra.11}
\eneq
\noindent
Eq.(\ref{ftra.11}) also implies 
\beq
 G_{+ , -}^{(\nu) }  ( \omega ) = \frac{l_c^\nu}{v^\nu\Gamma (  \nu) }
 \left( \frac{\hbar\beta }{2 \pi  } \right)^{1 - \nu } 
 e^{  -\frac{\hbar \omega \beta}{2} }  \left| \Gamma 
 \left( \frac{\nu }{2}  +  i \frac{\hbar \omega \beta}{2 \pi} \right)  \right|^2
 \, . 
  \label{ftra.12}
\eneq
\noindent
Following exactly the
same strategy of splitting the integral over $t$ into an integral 
from $- \infty$ to 0 plus and integral from $0 $ to $\infty$ and 
separately manipulating the two integrals as we have done before, 
one eventually finds 
 
\begin{multline}
 G_{+, +}^{(\nu) } ( \omega ) = 
  \frac{ l_c^\nu  e^{ -  i \pi \nu  /2 } }{v^\nu \Gamma (  \nu  )  \cos \left( \frac{\pi \nu }{2} \right)}
  \left( \frac{\hbar \beta}{2 \pi} \right)^{1-  \nu }  
  \\ \times
  \cosh \left( \frac{\hbar \omega \beta}{2} \right)  
 \left| \Gamma \left( \frac{ \nu  }{2}  +  i 
 \frac{\hbar \omega \beta}{2 \pi} \right)  \right|^2 \, ,
 \label{fad.1}
 \end{multline}
 \noindent
and 
\begin{multline}
 G_{- , -}^{(\nu) } ( \omega ) =   
 \frac{ l_c^\nu  e^{  i \pi \nu  /2 } }{v^\nu \Gamma (  \nu  )  \cos \left( \frac{\pi \nu }{2} \right)} 
 \left( \frac{\hbar \beta}{2 \pi} \right)^{1-  \nu }  
 \\ \times
 \cosh \left( \frac{\hbar \omega \beta}{2} \right)  
 \left| \Gamma \left( \frac{ \nu  }{2}  +  i \frac{\hbar \omega \beta}{2 \pi} \right)  
 \right|^2 \, .
 \label{fad.2}
 \end{multline}
 \noindent
 Eqs.(\ref{ftra.11},\ref{ftra.12},\ref{fad.1},\ref{fad.2}) provide us with 
 the Fourier transforms of the Keldysh Green functions which we used
 in the main text to compute the current correlations. Additional
 function one needs in performing the calculation are the  
Fourier transform of  $\xi_\pm ( t )$. These are given by  
\begin{multline}
\xi_\eta ( \omega ) = \int_{ - \infty}^\infty \: d t \: 
e^{ i \omega t } \xi_\eta ( t )  \\ =
i \int_{ - \infty}^\infty \: d u  \: e^{ i \frac{\hbar \omega\beta }{\pi} u } 
\left\{ \frac{e^{  u } e^{ -  i \eta \frac{\pi \tau_c}{\hbar \beta} } + 
e^{ -   u } e^{   i \eta \frac{\pi \tau_c}{\hbar \beta} } }{e^{  u } 
e^{ -  i \eta \frac{\pi \tau_c}{\hbar \beta} } - 
e^{ -   u } e^{   i \eta \frac{\pi \tau_c}{\hbar \beta} }} \right\}
\,, 
\label{cor.1a}
\end{multline}
\noindent
where we have set $u \equiv t \pi / \hbar \beta$. When  $\omega \neq 0$, 
a straightforward application of residue theorem gives
\begin{eqnarray}
 \xi_+ ( \omega ) &=&    \int_{- \infty}^\infty \: d \omega  
 \: e^{ i \omega t } \xi_+ ( t ) = \frac{2 \pi e^{ - \omega \tau_c} 
}{1 - e^{ - \hbar\beta \omega}}\, ,  \nonumber \\
 \xi_- ( \omega ) &=&    \int_{- \infty}^\infty \: d \omega  
 \: e^{ i \omega t } \xi_- ( t ) =\frac{2\pi e^{ \omega \tau_c} }{e^{ \hbar \beta \omega} - 1}
\,.
\label{cor.a2}
\end{eqnarray}
\noindent
The integrals in Eq.(\ref{cor.1a}) are ill-defined if $\omega = 0 $, this motivates 
the need of first computing the current current correlation in Fourier space at finite 
frequency $\Omega$, and then only afterwards taking $\Omega \to 0$. 

\section{Current-current correlation for chiral fermions.}
\label{chifer}

In this Appendix to check the consistency of the formulas we derived 
in section \ref{curcorre} with the standard results obtained by   B\"uttiker 
in the non-interacting case, we derive the current-current  
correlation in the case  Integer Quantum Hall
(IQH)-effect at filling  $\nu = 1$.  At $\nu = 1$, IQH-edges $e_1$ and $e_2$ (see Fig. \ref{setup1}) are 
described by the non-interacting chiral fermion  Hamiltonian:
\begin{align}
H_0=-i \hbar v\sum_{j=1}^2\int dx :\psi_j^\dag(x) \partial_x \psi_j(x): 
\;\;\;\; ,
\label{cfham}
\end{align}
\noindent
with $v$ being the Fermi velocity.
In the momentum basis, the chiral fermionic fields $\psi_i ( x ) $ take the 
mode-expansion 
\begin{equation}
\psi_j(x)=\frac{1}{\sqrt{\mathcal{L}_j}}\sum_{k_j}e^{ik_j x}c_{k_j,j} \, .
\label{electron_field_operator}
\end{equation}
\noindent
In Eq.(\ref{electron_field_operator}), we use  $\mathcal{L}_i$ to denote the length of the edge $i$, 
which we assume to be  large enough to be irrelevant for our final result.  The 
double columns $: \: :$ denote normal ordering with  respect to the groundstate 
$| GS \rangle = \prod_{i=1,2; \epsilon(k_j)\leq 0} c_{k_j,i}^\dag |0\rangle$.  $c_{k_j,i}$ 
is the electron annihilation operator  for a state with momentum  $k_j$ on edge $i$. 
Creation and annihilation operators in the momentum basis  satisfy the standard fermionic anti-commutations rules,
$\{c_{k_j,j}^\dag,c_{k_{j'}',j'}\}=\delta_{k_j,k_{j'}'}\delta_{j,j'}$,  
 $\{c_{k_j,j},c_{k_{j'}',j'}\}=0$. To account for the chemical potential 
 bias between the edges, we assume that, in absence of tunnelling, 
each edge $i$ is at equilibrium with a reservoir at chemical potential $\mu_i$. 

In order to allow for electrons to tunnel between the two edges, we consider the 
tunneling Hamiltonian $H_T$ given by  
\begin{equation}
H_T=\Gamma_e \psi^\dag_1(0)\psi_2(0)+{\rm h.c. }
\:\:\:\: .
\label{tunhamf}
\end{equation}
\noindent
The current density operator at site $x$ of edge $i$ is given by $i_i(x)=e v:\psi^\dag(x) \psi(x):$.
The current correlation function  between point $x_1$ 
on $e_1$  and point $x_2$ on $e_2$   (cfr Fig.~\ref{setup1})  is defined as in Eq.(\ref{correlation-anyons}) and, 
resorting again to the Keldysh formalism we use in section \ref{curcorre}, we readily find that  the 
corresponding zero-frequency limit, $S ( 0 )$, is  given by

\begin{multline}
S(0)= \frac{e^2 v^2}{2} \sum_{\eta=\pm 1}\int_{-\infty}^{+\infty}d(t_1-t_2)  \\
\langle \langle T_K :\hat{\psi}_1^\dag(x_1,t_1+\eta\, 0^+, \eta) \hat{\psi}_1(x_1,t_1, \eta): \\
:\hat{\psi}_2^\dag(x_2,t_2-\eta\, 0^+, -\eta) \hat{\psi}_2(x_2,t_2,- \eta):
\rangle \rangle\: . 
\label{curcof}
\end{multline}
\noindent
(note that, in order to preserve the correct ordering of 
the fermionic operators under the action of $T_K$, in Eq.(\ref{curcof})
we introduced  the infinitesimal positive quantity $0^+$ as a regularizator). 
Assuming a weak tunneling rate between the edges to recover consistency
with the analysis of  section \ref{curcorre}, we compute 
$S ( 0 )$ to second order in $H_T$, obtaining 
\begin{widetext}
\begin{multline}
S(0)=-\frac{|\Gamma_e|^2 e^2 v^2}{4 \hbar^2} \sum_{\eta,\eta_1,\eta_2=\pm1}\eta_1\eta_2 
\int_{-\infty}^{+\infty}d(t_1-t_2) \int_{-\infty}^{+\infty}ds_1 \int_{-\infty}^{+\infty}ds_2 \\
\times \left[
\mathcal{G}^{(1)}_{\eta_2,\eta}(-x_1,s_2-t_1)\mathcal{G}^{(1)}_{\eta,\eta_1}(x_1,t_1-s_1) 
\mathcal{G}^{(2)}_{\eta_1,-\eta}(-x_2,s_1-t_2)\mathcal{G}^{(2)}_{-\eta,\eta_2}(x_2,t_2-s_2)
\right. \\+\left.
\mathcal{G}^{(1)}_{\eta_1,\eta}(-x_1,s_1-t_1)\mathcal{G}^{(1)}_{\eta,\eta_2}(x_1,t_1-s_2) 
\mathcal{G}^{(2)}_{\eta_2,-\eta}(-x_2,s_2-t_2)\mathcal{G}^{(2)}_{-\eta,\eta_1}(x_2,t_2-s_1)
\right],
\label{gresfert}
\end{multline}
\end{widetext}
\noindent
with  the fermionic  Keldysh Green function 
$\mathcal{G}^{(i)}_{\eta_1,\eta_2}(x_1-x_2,t_1-t_2) = 
-i \langle T_K \psi_{i } (x_1,t_1 , \eta_1 ) \psi^\dag_{i  } (x_2,t_2 , \eta_2)\rangle$,
with $\psi_{ i } ( x , t ) $ being the fermion fields in the interaction representation with 
respect the Hamiltonian $H_0$. Moving to Fourier space, we may rewrite Eq.(\ref{gresfert})
as 
 
\begin{multline}
S(0)=\frac{|\Gamma_e|^2e^2 v^2}{ 2 \hbar^2} \sum_{\eta,\eta_1,\eta_2=\pm1}\eta_1\eta_2 
\int_{-\infty}^{+\infty}\frac{d\omega}{2 \pi} \\ \times \left[
\mathcal{G}^{(1)}_{\eta_2,\eta}(-x_1,\omega)\mathcal{G}^{(1)}_{\eta,\eta_1}(x_1,\omega) 
\mathcal{G}^{(2)}_{\eta_1,-\eta}(-x_2,\omega)\mathcal{G}^{(2)}_{-\eta,\eta_2}(x_2,\omega)
\right],
\label{gresferome}
\end{multline}
\noindent
with the singe-fermion Keldysh Green functions in Fourier space given by\cite{Campagnano:2013}
 \begin{align} 
\mathcal{G}^{(i)}_{++}( x,\omega)=&\frac{i}{v}e^{i \omega  x/ v}[f(\hbar \omega-\mu_i)-\Theta( x)]  \, , 
\\ 
\mathcal{G}^{(i)}_{+-}( x,\omega)=&\frac{i}{v}e^{i \omega  x/v} f(\hbar \omega-\mu_i) \, , \\  
\mathcal{G}^{(i)}_{-+}( x,\omega)=&-\frac{i}{v}e^{i \omega  x/ v} [1-f(\hbar \omega-\mu_i)]  \, , \\ 
\label{FT4}
\mathcal{G}^{(i)}_{--}( x,\omega)=&\frac{i}{v}e^{i \omega  x/ v}[f(\hbar \omega-\mu_i)-\Theta(- x)]  \, .
\end{align}
\noindent
In Eqs.(\ref{FT4})  $f(\omega)$ denotes the Fermi-Dirac distribution function 
$f ( \omega ) = [1+\exp(\beta \hbar \omega)]^{-1}$, while $\Theta ( x )$ is  the Heaviside step function
regularized so that  $\Theta(0)=1/2$. 
 
Performing the sum over the Keldysh indices, using Eqs.(\ref{FT4}) for 
both $x_1$ and $x_2 > 0$ , we obtain  
\begin{equation}
S(0)=-\frac{e^2|\Gamma_e|^2}{ 2\pi v^2 \hbar^2}\int_{-\infty}^{+\infty} 
d\omega \left[  f(\hbar\omega-\mu_1)-f(\hbar\omega-\mu_2)\right]^2.
\label{curcof.1}
\end{equation}
\noindent
Notice that in Eq.(\ref{tunham}) the tunneling amplitude $\Gamma$ has the dimension of an energy, while $\Gamma_e$ has the dimensions of an energy times a length.  
Eq.(\ref{curr-curr3}) evaluated for $\nu=1$ reproduce Eq.(\ref{curcof.1}) by taking
$\Gamma=\Gamma_e/(2 \pi l_c )$, which is indeed consistent with the bosonization identity\cite{Vondelft:1998}
$\psi_i(x)=e^{- i \phi_i(x)}/\sqrt{2\pi l_c}$ .


\begin{thebibliography}{59}
\expandafter\ifx\csname natexlab\endcsname\relax\def\natexlab#1{#1}\fi
\expandafter\ifx\csname bibnamefont\endcsname\relax
  \def\bibnamefont#1{#1}\fi
\expandafter\ifx\csname bibfnamefont\endcsname\relax
  \def\bibfnamefont#1{#1}\fi
\expandafter\ifx\csname citenamefont\endcsname\relax
  \def\citenamefont#1{#1}\fi
\expandafter\ifx\csname url\endcsname\relax
  \def\url#1{\texttt{#1}}\fi
\expandafter\ifx\csname urlprefix\endcsname\relax\def\urlprefix{URL }\fi
\providecommand{\bibinfo}[2]{#2}
\providecommand{\eprint}[2][]{\url{#2}}

\bibitem[{\citenamefont{B\"uttiker}(1992)}]{Buettiker:1992}
\bibinfo{author}{\bibfnamefont{M.}~\bibnamefont{B\"uttiker}},
  \bibinfo{journal}{Phys. Rev. B} \textbf{\bibinfo{volume}{46}},
  \bibinfo{pages}{12485} (\bibinfo{year}{1992}).

\bibitem[{\citenamefont{Tsui et~al.}(1982)\citenamefont{Tsui, Stormer, and
  Gossard}}]{tsui}
\bibinfo{author}{\bibfnamefont{D.~C.} \bibnamefont{Tsui}},
  \bibinfo{author}{\bibfnamefont{H.~L.} \bibnamefont{Stormer}},
  \bibnamefont{and} \bibinfo{author}{\bibfnamefont{A.~C.}
  \bibnamefont{Gossard}}, \bibinfo{journal}{Phys. Rev. Lett.}
  \textbf{\bibinfo{volume}{48}}, \bibinfo{pages}{1559} (\bibinfo{year}{1982}).

\bibitem[{\citenamefont{Laughlin}(1983)}]{laughlin}
\bibinfo{author}{\bibfnamefont{R.~B.} \bibnamefont{Laughlin}},
  \bibinfo{journal}{Phys. Rev. Lett.} \textbf{\bibinfo{volume}{50}},
  \bibinfo{pages}{1395} (\bibinfo{year}{1983}).

\bibitem[{\citenamefont{Arovas et~al.}(1984)\citenamefont{Arovas, Schrieffer,
  and Wilczek}}]{Arovas:1984}
\bibinfo{author}{\bibfnamefont{D.}~\bibnamefont{Arovas}},
  \bibinfo{author}{\bibfnamefont{J.~R.} \bibnamefont{Schrieffer}},
  \bibnamefont{and} \bibinfo{author}{\bibfnamefont{F.}~\bibnamefont{Wilczek}},
  \bibinfo{journal}{Phys. Rev. Lett.} \textbf{\bibinfo{volume}{53}},
  \bibinfo{pages}{722} (\bibinfo{year}{1984}).

\bibitem[{\citenamefont{Nayak et~al.}(2008)\citenamefont{Nayak, Simon, Stern,
  Freedman, and Das~Sarma}}]{Najak:2008}
\bibinfo{author}{\bibfnamefont{C.}~\bibnamefont{Nayak}},
  \bibinfo{author}{\bibfnamefont{S.~H.} \bibnamefont{Simon}},
  \bibinfo{author}{\bibfnamefont{A.}~\bibnamefont{Stern}},
  \bibinfo{author}{\bibfnamefont{M.}~\bibnamefont{Freedman}}, \bibnamefont{and}
  \bibinfo{author}{\bibfnamefont{S.}~\bibnamefont{Das~Sarma}},
  \bibinfo{journal}{Rev. Mod. Phys.} \textbf{\bibinfo{volume}{80}},
  \bibinfo{pages}{1083} (\bibinfo{year}{2008}).

\bibitem[{\citenamefont{Wen}(2004)}]{wenbook}
\bibinfo{author}{\bibfnamefont{X.~G.} \bibnamefont{Wen}},
  \emph{\bibinfo{title}{Quantum Field Theory Of Many-Body Systems: From The
  Origin Of Sound To An Origin Of Light And Electrons}}
  (\bibinfo{publisher}{Oxford University Press}, \bibinfo{year}{2004}).

\bibitem[{\citenamefont{de~Picciotto et~al.}(1997)\citenamefont{de~Picciotto,
  Reznikov, Heiblum, Umansky, Bunin, and Mahalu}}]{de-Picciotto:1997}
\bibinfo{author}{\bibfnamefont{R.}~\bibnamefont{de~Picciotto}},
  \bibinfo{author}{\bibfnamefont{M.}~\bibnamefont{Reznikov}},
  \bibinfo{author}{\bibfnamefont{M.}~\bibnamefont{Heiblum}},
  \bibinfo{author}{\bibfnamefont{V.}~\bibnamefont{Umansky}},
  \bibinfo{author}{\bibfnamefont{G.}~\bibnamefont{Bunin}}, \bibnamefont{and}
  \bibinfo{author}{\bibfnamefont{D.}~\bibnamefont{Mahalu}},
  \bibinfo{journal}{Nature} \textbf{\bibinfo{volume}{389}},
  \bibinfo{pages}{162} (\bibinfo{year}{1997}).

\bibitem[{\citenamefont{Saminadayar et~al.}(1997)\citenamefont{Saminadayar,
  Glattli, Jin, and Etienne}}]{Saminadayar:1997}
\bibinfo{author}{\bibfnamefont{L.}~\bibnamefont{Saminadayar}},
  \bibinfo{author}{\bibfnamefont{D.~C.} \bibnamefont{Glattli}},
  \bibinfo{author}{\bibfnamefont{Y.}~\bibnamefont{Jin}}, \bibnamefont{and}
  \bibinfo{author}{\bibfnamefont{B.}~\bibnamefont{Etienne}},
  \bibinfo{journal}{Phys. Rev. Lett.} \textbf{\bibinfo{volume}{79}},
  \bibinfo{pages}{2526} (\bibinfo{year}{1997}).

\bibitem[{\citenamefont{Camino et~al.}(2005)\citenamefont{Camino, Zhou, and
  Goldman}}]{Camino:2005}
\bibinfo{author}{\bibfnamefont{F.~E.} \bibnamefont{Camino}},
  \bibinfo{author}{\bibfnamefont{W.}~\bibnamefont{Zhou}}, \bibnamefont{and}
  \bibinfo{author}{\bibfnamefont{V.~J.} \bibnamefont{Goldman}},
  \bibinfo{journal}{Phys. Rev. B} \textbf{\bibinfo{volume}{72}},
  \bibinfo{pages}{075342} (\bibinfo{year}{2005}).

\bibitem[{\citenamefont{Willett et~al.}(2013)\citenamefont{Willett, Nayak,
  Shtengel, Pfeiffer, and West}}]{Willett:2013}
\bibinfo{author}{\bibfnamefont{R.~L.} \bibnamefont{Willett}},
  \bibinfo{author}{\bibfnamefont{C.}~\bibnamefont{Nayak}},
  \bibinfo{author}{\bibfnamefont{K.}~\bibnamefont{Shtengel}},
  \bibinfo{author}{\bibfnamefont{L.~N.} \bibnamefont{Pfeiffer}},
  \bibnamefont{and} \bibinfo{author}{\bibfnamefont{K.~W.} \bibnamefont{West}},
  \bibinfo{journal}{Phys. Rev. Lett.} \textbf{\bibinfo{volume}{111}},
  \bibinfo{pages}{186401} (\bibinfo{year}{2013}).

\bibitem[{\citenamefont{Hanbury~Brown and Twiss}(1954)}]{Hanbury:1954}
\bibinfo{author}{\bibfnamefont{R.}~\bibnamefont{Hanbury~Brown}}
  \bibnamefont{and} \bibinfo{author}{\bibfnamefont{R.~Q.} \bibnamefont{Twiss}},
  \bibinfo{journal}{Phil. Mag.} \textbf{\bibinfo{volume}{45}},
  \bibinfo{pages}{663} (\bibinfo{year}{1954}).

\bibitem[{\citenamefont{Hanbury~Brown and Twiss}(1956)}]{Hanbury:1955}
\bibinfo{author}{\bibfnamefont{R.}~\bibnamefont{Hanbury~Brown}}
  \bibnamefont{and} \bibinfo{author}{\bibfnamefont{R.~Q.} \bibnamefont{Twiss}},
  \bibinfo{journal}{Phys. Rev. B} \textbf{\bibinfo{volume}{177}},
  \bibinfo{pages}{27} (\bibinfo{year}{1956}).

\bibitem[{\citenamefont{Oliver et~al.}(1999)\citenamefont{Oliver, Kim, Liu, and
  Yamamoto}}]{Oliver:1999}
\bibinfo{author}{\bibfnamefont{W.~D.} \bibnamefont{Oliver}},
  \bibinfo{author}{\bibfnamefont{J.}~\bibnamefont{Kim}},
  \bibinfo{author}{\bibfnamefont{R.~C.} \bibnamefont{Liu}}, \bibnamefont{and}
  \bibinfo{author}{\bibfnamefont{Y.}~\bibnamefont{Yamamoto}},
  \bibinfo{journal}{Science} \textbf{\bibinfo{volume}{284}},
  \bibinfo{pages}{299} (\bibinfo{year}{1999}).

\bibitem[{\citenamefont{de~C.~Chamon et~al.}(1997)\citenamefont{de~C.~Chamon,
  Freed, Kivelson, Sondhi, and Wen}}]{Chamon:1997}
\bibinfo{author}{\bibfnamefont{C.}~\bibnamefont{de~C.~Chamon}},
  \bibinfo{author}{\bibfnamefont{D.~E.} \bibnamefont{Freed}},
  \bibinfo{author}{\bibfnamefont{S.~A.} \bibnamefont{Kivelson}},
  \bibinfo{author}{\bibfnamefont{S.~L.} \bibnamefont{Sondhi}},
  \bibnamefont{and} \bibinfo{author}{\bibfnamefont{X.~G.} \bibnamefont{Wen}},
  \bibinfo{journal}{Phys. Rev. B} \textbf{\bibinfo{volume}{55}},
  \bibinfo{pages}{2331} (\bibinfo{year}{1997}).

\bibitem[{\citenamefont{Stern and Halperin}(2006)}]{Stern:2006}
\bibinfo{author}{\bibfnamefont{A.}~\bibnamefont{Stern}} \bibnamefont{and}
  \bibinfo{author}{\bibfnamefont{B.~I.} \bibnamefont{Halperin}},
  \bibinfo{journal}{Phys. Rev. Lett.} \textbf{\bibinfo{volume}{96}},
  \bibinfo{pages}{016802} (\bibinfo{year}{2006}).

\bibitem[{\citenamefont{Bonderson
  et~al.}(2006{\natexlab{a}})\citenamefont{Bonderson, Kitaev, and
  Shtengel}}]{Bonderson:2006}
\bibinfo{author}{\bibfnamefont{P.}~\bibnamefont{Bonderson}},
  \bibinfo{author}{\bibfnamefont{A.}~\bibnamefont{Kitaev}}, \bibnamefont{and}
  \bibinfo{author}{\bibfnamefont{K.}~\bibnamefont{Shtengel}},
  \bibinfo{journal}{Phys. Rev. Lett.} \textbf{\bibinfo{volume}{96}},
  \bibinfo{pages}{016803} (\bibinfo{year}{2006}{\natexlab{a}}).

\bibitem[{\citenamefont{Bonderson
  et~al.}(2006{\natexlab{b}})\citenamefont{Bonderson, Shtengel, and
  Slingerland}}]{Bonderson:2006b}
\bibinfo{author}{\bibfnamefont{P.}~\bibnamefont{Bonderson}},
  \bibinfo{author}{\bibfnamefont{K.}~\bibnamefont{Shtengel}}, \bibnamefont{and}
  \bibinfo{author}{\bibfnamefont{J.~K.} \bibnamefont{Slingerland}},
  \bibinfo{journal}{Phys. Rev. Lett.} \textbf{\bibinfo{volume}{97}},
  \bibinfo{pages}{016401} (\bibinfo{year}{2006}{\natexlab{b}}).

\bibitem[{\citenamefont{Ardonne and Kim}(2008)}]{Kim:2008}
\bibinfo{author}{\bibfnamefont{E.}~\bibnamefont{Ardonne}} \bibnamefont{and}
  \bibinfo{author}{\bibfnamefont{E.-A.} \bibnamefont{Kim}},
  \bibinfo{journal}{Journal of Statistical Mechanics: Theory and Experiment}
  \textbf{\bibinfo{volume}{2008}}, \bibinfo{pages}{L04001}
  (\bibinfo{year}{2008}).

\bibitem[{\citenamefont{Bishara et~al.}(2009)\citenamefont{Bishara, Bonderson,
  Nayak, Shtengel, and Slingerland}}]{Bishara:2009}
\bibinfo{author}{\bibfnamefont{W.}~\bibnamefont{Bishara}},
  \bibinfo{author}{\bibfnamefont{P.}~\bibnamefont{Bonderson}},
  \bibinfo{author}{\bibfnamefont{C.}~\bibnamefont{Nayak}},
  \bibinfo{author}{\bibfnamefont{K.}~\bibnamefont{Shtengel}}, \bibnamefont{and}
  \bibinfo{author}{\bibfnamefont{J.~K.} \bibnamefont{Slingerland}},
  \bibinfo{journal}{Phys. Rev. B} \textbf{\bibinfo{volume}{80}},
  \bibinfo{pages}{155303} (\bibinfo{year}{2009}).

\bibitem[{\citenamefont{Halperin et~al.}(2011)\citenamefont{Halperin, Stern,
  Neder, and Rosenow}}]{Halperin:2011}
\bibinfo{author}{\bibfnamefont{B.~I.} \bibnamefont{Halperin}},
  \bibinfo{author}{\bibfnamefont{A.}~\bibnamefont{Stern}},
  \bibinfo{author}{\bibfnamefont{I.}~\bibnamefont{Neder}}, \bibnamefont{and}
  \bibinfo{author}{\bibfnamefont{B.}~\bibnamefont{Rosenow}},
  \bibinfo{journal}{Phys. Rev. B} \textbf{\bibinfo{volume}{83}},
  \bibinfo{pages}{155440} (\bibinfo{year}{2011}).

\bibitem[{\citenamefont{Law et~al.}(2006)\citenamefont{Law, Feldman, and
  Gefen}}]{Feldman:2006}
\bibinfo{author}{\bibfnamefont{K.~T.} \bibnamefont{Law}},
  \bibinfo{author}{\bibfnamefont{D.~E.} \bibnamefont{Feldman}},
  \bibnamefont{and} \bibinfo{author}{\bibfnamefont{Y.}~\bibnamefont{Gefen}},
  \bibinfo{journal}{Phys. Rev. B} \textbf{\bibinfo{volume}{74}},
  \bibinfo{pages}{045319} (\bibinfo{year}{2006}).

\bibitem[{\citenamefont{Feldman et~al.}(2007)\citenamefont{Feldman, Gefen,
  Kitaev, Law, and Stern}}]{Feldman:2007}
\bibinfo{author}{\bibfnamefont{D.~E.} \bibnamefont{Feldman}},
  \bibinfo{author}{\bibfnamefont{Y.}~\bibnamefont{Gefen}},
  \bibinfo{author}{\bibfnamefont{A.}~\bibnamefont{Kitaev}},
  \bibinfo{author}{\bibfnamefont{K.~T.} \bibnamefont{Law}}, \bibnamefont{and}
  \bibinfo{author}{\bibfnamefont{A.}~\bibnamefont{Stern}},
  \bibinfo{journal}{Phys. Rev. B} \textbf{\bibinfo{volume}{76}},
  \bibinfo{pages}{085333} (\bibinfo{year}{2007}).

\bibitem[{\citenamefont{Ponomarenko and Averin}(2007)}]{Ponomarenko:2007}
\bibinfo{author}{\bibfnamefont{V.~V.} \bibnamefont{Ponomarenko}}
  \bibnamefont{and} \bibinfo{author}{\bibfnamefont{D.~V.}
  \bibnamefont{Averin}}, \bibinfo{journal}{Phys. Rev. Lett.}
  \textbf{\bibinfo{volume}{99}}, \bibinfo{pages}{066803}
  (\bibinfo{year}{2007}).

\bibitem[{\citenamefont{Ponomarenko and Averin}(2009)}]{Ponomarenko:2009}
\bibinfo{author}{\bibfnamefont{V.~V.} \bibnamefont{Ponomarenko}}
  \bibnamefont{and} \bibinfo{author}{\bibfnamefont{D.~V.}
  \bibnamefont{Averin}}, \bibinfo{journal}{Phys. Rev. B}
  \textbf{\bibinfo{volume}{79}}, \bibinfo{pages}{045303}
  (\bibinfo{year}{2009}).

\bibitem[{\citenamefont{Bonderson et~al.}(2008)\citenamefont{Bonderson,
  Shtengel, and Slingerland}}]{Bonderson:2008}
\bibinfo{author}{\bibfnamefont{P.}~\bibnamefont{Bonderson}},
  \bibinfo{author}{\bibfnamefont{K.}~\bibnamefont{Shtengel}}, \bibnamefont{and}
  \bibinfo{author}{\bibfnamefont{J.}~\bibnamefont{Slingerland}},
  \bibinfo{journal}{Annals of Physics} \textbf{\bibinfo{volume}{323}},
  \bibinfo{pages}{2709 } (\bibinfo{year}{2008}), ISSN
  \bibinfo{issn}{0003-4916}.

\bibitem[{\citenamefont{Law}(2008)}]{Law:2008}
\bibinfo{author}{\bibfnamefont{K.~T.} \bibnamefont{Law}},
  \bibinfo{journal}{Phys. Rev. B} \textbf{\bibinfo{volume}{77}},
  \bibinfo{pages}{205310} (\bibinfo{year}{2008}).

\bibitem[{\citenamefont{Levkivskyi et~al.}(2009)\citenamefont{Levkivskyi,
  Boyarsky, Fr\"ohlich, and Sukhorukov}}]{Levkivskyi:2009}
\bibinfo{author}{\bibfnamefont{I.~P.} \bibnamefont{Levkivskyi}},
  \bibinfo{author}{\bibfnamefont{A.}~\bibnamefont{Boyarsky}},
  \bibinfo{author}{\bibfnamefont{J.}~\bibnamefont{Fr\"ohlich}},
  \bibnamefont{and} \bibinfo{author}{\bibfnamefont{E.~V.}
  \bibnamefont{Sukhorukov}}, \bibinfo{journal}{Phys. Rev. B}
  \textbf{\bibinfo{volume}{80}}, \bibinfo{pages}{045319}
  (\bibinfo{year}{2009}).

\bibitem[{\citenamefont{Ponomarenko and Averin}(2010)}]{Ponomarenko:2010}
\bibinfo{author}{\bibfnamefont{V.~V.} \bibnamefont{Ponomarenko}}
  \bibnamefont{and} \bibinfo{author}{\bibfnamefont{D.~V.}
  \bibnamefont{Averin}}, \bibinfo{journal}{Phys. Rev. B}
  \textbf{\bibinfo{volume}{82}}, \bibinfo{pages}{205411}
  (\bibinfo{year}{2010}).

\bibitem[{\citenamefont{Wang and Feldman}(2010)}]{Wang:2010}
\bibinfo{author}{\bibfnamefont{C.}~\bibnamefont{Wang}} \bibnamefont{and}
  \bibinfo{author}{\bibfnamefont{D.~E.} \bibnamefont{Feldman}},
  \bibinfo{journal}{Phys. Rev. B} \textbf{\bibinfo{volume}{82}},
  \bibinfo{pages}{165314} (\bibinfo{year}{2010}).

\bibitem[{\citenamefont{Levkivskyi et~al.}(2012)\citenamefont{Levkivskyi,
  Fr\"ohlich, and Sukhorukov}}]{Levkivskyi:2012}
\bibinfo{author}{\bibfnamefont{I.~P.} \bibnamefont{Levkivskyi}},
  \bibinfo{author}{\bibfnamefont{J.}~\bibnamefont{Fr\"ohlich}},
  \bibnamefont{and} \bibinfo{author}{\bibfnamefont{E.~V.}
  \bibnamefont{Sukhorukov}}, \bibinfo{journal}{Phys. Rev. B}
  \textbf{\bibinfo{volume}{86}}, \bibinfo{pages}{245105}
  (\bibinfo{year}{2012}).

\bibitem[{\citenamefont{Ganeshan et~al.}(2012)\citenamefont{Ganeshan, Abanov,
  and Averin}}]{Ganeshan:2012}
\bibinfo{author}{\bibfnamefont{S.}~\bibnamefont{Ganeshan}},
  \bibinfo{author}{\bibfnamefont{A.~G.} \bibnamefont{Abanov}},
  \bibnamefont{and} \bibinfo{author}{\bibfnamefont{D.~V.}
  \bibnamefont{Averin}}, \bibinfo{journal}{Phys. Rev. B}
  \textbf{\bibinfo{volume}{86}}, \bibinfo{pages}{235309}
  (\bibinfo{year}{2012}).

\bibitem[{\citenamefont{Yang}(2015)}]{Yang:2015}
\bibinfo{author}{\bibfnamefont{G.}~\bibnamefont{Yang}}, \bibinfo{journal}{Phys.
  Rev. B} \textbf{\bibinfo{volume}{91}}, \bibinfo{pages}{115109}
  (\bibinfo{year}{2015}).

\bibitem[{\citenamefont{Safi et~al.}(2001)\citenamefont{Safi, Devillard, and
  Martin}}]{Safi:2001}
\bibinfo{author}{\bibfnamefont{I.}~\bibnamefont{Safi}},
  \bibinfo{author}{\bibfnamefont{P.}~\bibnamefont{Devillard}},
  \bibnamefont{and} \bibinfo{author}{\bibfnamefont{T.}~\bibnamefont{Martin}},
  \bibinfo{journal}{Phys. Rev. Lett.} \textbf{\bibinfo{volume}{86}},
  \bibinfo{pages}{4628} (\bibinfo{year}{2001}).

\bibitem[{\citenamefont{Vishveshwara}(2003)}]{Vishveshwara:2003}
\bibinfo{author}{\bibfnamefont{S.}~\bibnamefont{Vishveshwara}},
  \bibinfo{journal}{Phys. Rev. Lett.} \textbf{\bibinfo{volume}{91}},
  \bibinfo{pages}{196803} (\bibinfo{year}{2003}).

\bibitem[{\citenamefont{Kim et~al.}(2005)\citenamefont{Kim, Lawler,
  Vishveshwara, and Fradkin}}]{Kim:2005}
\bibinfo{author}{\bibfnamefont{E.-A.} \bibnamefont{Kim}},
  \bibinfo{author}{\bibfnamefont{M.}~\bibnamefont{Lawler}},
  \bibinfo{author}{\bibfnamefont{S.}~\bibnamefont{Vishveshwara}},
  \bibnamefont{and} \bibinfo{author}{\bibfnamefont{E.}~\bibnamefont{Fradkin}},
  \bibinfo{journal}{Phys. Rev. Lett.} \textbf{\bibinfo{volume}{95}},
  \bibinfo{pages}{176402} (\bibinfo{year}{2005}).

\bibitem[{\citenamefont{Kim et~al.}(2006)\citenamefont{Kim, Lawler,
  Vishveshwara, and Fradkin}}]{Kim:2006}
\bibinfo{author}{\bibfnamefont{E.-A.} \bibnamefont{Kim}},
  \bibinfo{author}{\bibfnamefont{M.~J.} \bibnamefont{Lawler}},
  \bibinfo{author}{\bibfnamefont{S.}~\bibnamefont{Vishveshwara}},
  \bibnamefont{and} \bibinfo{author}{\bibfnamefont{E.}~\bibnamefont{Fradkin}},
  \bibinfo{journal}{Phys. Rev. B} \textbf{\bibinfo{volume}{74}},
  \bibinfo{pages}{155324} (\bibinfo{year}{2006}).

\bibitem[{\citenamefont{Campagnano et~al.}(2012)\citenamefont{Campagnano,
  Zilberberg, Gornyi, Feldman, Potter, and Gefen}}]{Campagnano:2012}
\bibinfo{author}{\bibfnamefont{G.}~\bibnamefont{Campagnano}},
  \bibinfo{author}{\bibfnamefont{O.}~\bibnamefont{Zilberberg}},
  \bibinfo{author}{\bibfnamefont{I.~V.} \bibnamefont{Gornyi}},
  \bibinfo{author}{\bibfnamefont{D.~E.} \bibnamefont{Feldman}},
  \bibinfo{author}{\bibfnamefont{A.~C.} \bibnamefont{Potter}},
  \bibnamefont{and} \bibinfo{author}{\bibfnamefont{Y.}~\bibnamefont{Gefen}},
  \bibinfo{journal}{Phys. Rev. Lett.} \textbf{\bibinfo{volume}{109}},
  \bibinfo{pages}{106802} (\bibinfo{year}{2012}).

\bibitem[{\citenamefont{Campagnano et~al.}(2013)\citenamefont{Campagnano,
  Zilberberg, Gornyi, and Gefen}}]{Campagnano:2013}
\bibinfo{author}{\bibfnamefont{G.}~\bibnamefont{Campagnano}},
  \bibinfo{author}{\bibfnamefont{O.}~\bibnamefont{Zilberberg}},
  \bibinfo{author}{\bibfnamefont{I.~V.} \bibnamefont{Gornyi}},
  \bibnamefont{and} \bibinfo{author}{\bibfnamefont{Y.}~\bibnamefont{Gefen}},
  \bibinfo{journal}{Phys. Rev. B} \textbf{\bibinfo{volume}{88}},
  \bibinfo{pages}{235415} (\bibinfo{year}{2013}).

\bibitem[{\citenamefont{Blanter and Buettiker}(2000)}]{Blanter:2000}
\bibinfo{author}{\bibfnamefont{Y.}~\bibnamefont{Blanter}} \bibnamefont{and}
  \bibinfo{author}{\bibfnamefont{M.}~\bibnamefont{Buettiker}},
  \bibinfo{journal}{Physics Reports} \textbf{\bibinfo{volume}{336}},
  \bibinfo{pages}{1 } (\bibinfo{year}{2000}), ISSN \bibinfo{issn}{0370-1573}.

\bibitem[{\citenamefont{B\"uttiker}(1991)}]{Buettiker:1991}
\bibinfo{author}{\bibfnamefont{M.}~\bibnamefont{B\"uttiker}},
  \bibinfo{journal}{Physica B: Condensed Matter}
  \textbf{\bibinfo{volume}{175}}, \bibinfo{pages}{199 } (\bibinfo{year}{1991}),
  ISSN \bibinfo{issn}{0921-4526}, \bibinfo{note}{analogies in Optics and
  Micro-Electronics}.

\bibitem[{\citenamefont{Kane and Fisher}(1992)}]{Kane:1992}
\bibinfo{author}{\bibfnamefont{C.~L.} \bibnamefont{Kane}} \bibnamefont{and}
  \bibinfo{author}{\bibfnamefont{M.~P.~A.} \bibnamefont{Fisher}},
  \bibinfo{journal}{Phys. Rev. B} \textbf{\bibinfo{volume}{46}},
  \bibinfo{pages}{15233} (\bibinfo{year}{1992}).

\bibitem[{\citenamefont{Kane and Fisher}(1994)}]{Kane:1994b}
\bibinfo{author}{\bibfnamefont{C.~L.} \bibnamefont{Kane}} \bibnamefont{and}
  \bibinfo{author}{\bibfnamefont{M.~P.~A.} \bibnamefont{Fisher}},
  \bibinfo{journal}{Phys. Rev. Lett.} \textbf{\bibinfo{volume}{72}},
  \bibinfo{pages}{724} (\bibinfo{year}{1994}).

\bibitem[{\citenamefont{Guyon et~al.}(2002)\citenamefont{Guyon, Devillard,
  Martin, and Safi}}]{Guyon:2002}
\bibinfo{author}{\bibfnamefont{R.}~\bibnamefont{Guyon}},
  \bibinfo{author}{\bibfnamefont{P.}~\bibnamefont{Devillard}},
  \bibinfo{author}{\bibfnamefont{T.}~\bibnamefont{Martin}}, \bibnamefont{and}
  \bibinfo{author}{\bibfnamefont{I.}~\bibnamefont{Safi}},
  \bibinfo{journal}{Phys. Rev. B} \textbf{\bibinfo{volume}{65}},
  \bibinfo{pages}{153304} (\bibinfo{year}{2002}).

\bibitem[{\citenamefont{Lee et~al.}(2012)\citenamefont{Lee, Lee, and
  Chung}}]{Lee:2012}
\bibinfo{author}{\bibfnamefont{Y.-W.} \bibnamefont{Lee}},
  \bibinfo{author}{\bibfnamefont{Y.-L.} \bibnamefont{Lee}}, \bibnamefont{and}
  \bibinfo{author}{\bibfnamefont{C.-H.} \bibnamefont{Chung}},
  \bibinfo{journal}{Phys. Rev. B} \textbf{\bibinfo{volume}{86}},
  \bibinfo{pages}{235121} (\bibinfo{year}{2012}).

\bibitem[{\citenamefont{Wen}(1991)}]{Wen:1991}
\bibinfo{author}{\bibfnamefont{X.-G.} \bibnamefont{Wen}},
  \bibinfo{journal}{Phys. Rev. B} \textbf{\bibinfo{volume}{44}},
  \bibinfo{pages}{5708} (\bibinfo{year}{1991}).

\bibitem[{\citenamefont{Fendley
  et~al.}(1995{\natexlab{a}})\citenamefont{Fendley, Ludwig, and
  Saleur}}]{Fendley:1995}
\bibinfo{author}{\bibfnamefont{P.}~\bibnamefont{Fendley}},
  \bibinfo{author}{\bibfnamefont{A.~W.~W.} \bibnamefont{Ludwig}},
  \bibnamefont{and} \bibinfo{author}{\bibfnamefont{H.}~\bibnamefont{Saleur}},
  \bibinfo{journal}{Phys. Rev. Lett.} \textbf{\bibinfo{volume}{74}},
  \bibinfo{pages}{3005} (\bibinfo{year}{1995}{\natexlab{a}}).

\bibitem[{\citenamefont{Fendley
  et~al.}(1995{\natexlab{b}})\citenamefont{Fendley, Ludwig, and
  Saleur}}]{Fendley:1995c}
\bibinfo{author}{\bibfnamefont{P.}~\bibnamefont{Fendley}},
  \bibinfo{author}{\bibfnamefont{A.~W.~W.} \bibnamefont{Ludwig}},
  \bibnamefont{and} \bibinfo{author}{\bibfnamefont{H.}~\bibnamefont{Saleur}},
  \bibinfo{journal}{Phys. Rev. B} \textbf{\bibinfo{volume}{52}},
  \bibinfo{pages}{8934} (\bibinfo{year}{1995}{\natexlab{b}}).

\bibitem[{\citenamefont{Fendley and Saleur}(1996)}]{Fendley:1996}
\bibinfo{author}{\bibfnamefont{P.}~\bibnamefont{Fendley}} \bibnamefont{and}
  \bibinfo{author}{\bibfnamefont{H.}~\bibnamefont{Saleur}},
  \bibinfo{journal}{Phys. Rev. B} \textbf{\bibinfo{volume}{54}},
  \bibinfo{pages}{10845} (\bibinfo{year}{1996}).

\bibitem[{\citenamefont{Haldane}(1991)}]{Haldane:1991}
\bibinfo{author}{\bibfnamefont{F.~D.~M.} \bibnamefont{Haldane}},
  \bibinfo{journal}{Phys. Rev. Lett.} \textbf{\bibinfo{volume}{67}},
  \bibinfo{pages}{937} (\bibinfo{year}{1991}).

\bibitem[{\citenamefont{Rosenow et~al.}(2015)\citenamefont{Rosenow, Levkivskyi,
  and Halperin}}]{Rosenow:2015}
\bibinfo{author}{\bibfnamefont{B.}~\bibnamefont{Rosenow}},
  \bibinfo{author}{\bibfnamefont{I.}~\bibnamefont{Levkivskyi}},
  \bibnamefont{and} \bibinfo{author}{\bibfnamefont{B.}~\bibnamefont{Halperin}},
  \bibinfo{journal}{eprint arXiv:1509.08470}  (\bibinfo{year}{2015}).

\bibitem[{\citenamefont{Kane et~al.}(1994)\citenamefont{Kane, Fisher, and
  Polchinski}}]{Kane:1994}
\bibinfo{author}{\bibfnamefont{C.~L.} \bibnamefont{Kane}},
  \bibinfo{author}{\bibfnamefont{M.~P.~A.} \bibnamefont{Fisher}},
  \bibnamefont{and}
  \bibinfo{author}{\bibfnamefont{J.}~\bibnamefont{Polchinski}},
  \bibinfo{journal}{Phys. Rev. Lett.} \textbf{\bibinfo{volume}{72}},
  \bibinfo{pages}{4129} (\bibinfo{year}{1994}).

\bibitem[{\citenamefont{Kane and Fisher}(1995)}]{Kane:1995}
\bibinfo{author}{\bibfnamefont{C.~L.} \bibnamefont{Kane}} \bibnamefont{and}
  \bibinfo{author}{\bibfnamefont{M.~P.~A.} \bibnamefont{Fisher}},
  \bibinfo{journal}{Phys. Rev. B} \textbf{\bibinfo{volume}{51}},
  \bibinfo{pages}{13449} (\bibinfo{year}{1995}).

\bibitem[{\citenamefont{Bid et~al.}(2010)\citenamefont{Bid, Ofek, Inoue,
  Heiblum, Kane, Umansky, and Mahalu}}]{Bid:2010}
\bibinfo{author}{\bibfnamefont{A.}~\bibnamefont{Bid}},
  \bibinfo{author}{\bibfnamefont{N.}~\bibnamefont{Ofek}},
  \bibinfo{author}{\bibfnamefont{H.}~\bibnamefont{Inoue}},
  \bibinfo{author}{\bibfnamefont{M.}~\bibnamefont{Heiblum}},
  \bibinfo{author}{\bibfnamefont{C.~L.} \bibnamefont{Kane}},
  \bibinfo{author}{\bibfnamefont{V.}~\bibnamefont{Umansky}}, \bibnamefont{and}
  \bibinfo{author}{\bibfnamefont{D.}~\bibnamefont{Mahalu}},
  \bibinfo{journal}{Nature} \textbf{\bibinfo{volume}{466}},
  \bibinfo{pages}{585} (\bibinfo{year}{2010}).

\bibitem[{\citenamefont{Gross et~al.}(2012)\citenamefont{Gross, Dolev, Heiblum,
  Umansky, and Mahalu}}]{Yaron:2012}
\bibinfo{author}{\bibfnamefont{Y.}~\bibnamefont{Gross}},
  \bibinfo{author}{\bibfnamefont{M.}~\bibnamefont{Dolev}},
  \bibinfo{author}{\bibfnamefont{M.}~\bibnamefont{Heiblum}},
  \bibinfo{author}{\bibfnamefont{V.}~\bibnamefont{Umansky}}, \bibnamefont{and}
  \bibinfo{author}{\bibfnamefont{D.}~\bibnamefont{Mahalu}},
  \bibinfo{journal}{Phys. Rev. Lett.} \textbf{\bibinfo{volume}{108}},
  \bibinfo{pages}{226801} (\bibinfo{year}{2012}).

\bibitem[{\citenamefont{Gurman et~al.}(2012)\citenamefont{Gurman, Sabo,
  Heiblum, Umansky, and Mahalu}}]{Gurman:2012}
\bibinfo{author}{\bibfnamefont{I.}~\bibnamefont{Gurman}},
  \bibinfo{author}{\bibfnamefont{R.}~\bibnamefont{Sabo}},
  \bibinfo{author}{\bibfnamefont{M.}~\bibnamefont{Heiblum}},
  \bibinfo{author}{\bibfnamefont{V.}~\bibnamefont{Umansky}}, \bibnamefont{and}
  \bibinfo{author}{\bibfnamefont{D.}~\bibnamefont{Mahalu}},
  \bibinfo{journal}{Nature Comm.} \textbf{\bibinfo{volume}{3}},
  \bibinfo{pages}{1289} (\bibinfo{year}{2012}).

\bibitem[{\citenamefont{Bernevig
  et~al.}(2001{\natexlab{a}})\citenamefont{Bernevig, Giuliano, and
  Laughlin}}]{Giuliano:2001}
\bibinfo{author}{\bibfnamefont{B.~A.} \bibnamefont{Bernevig}},
  \bibinfo{author}{\bibfnamefont{D.}~\bibnamefont{Giuliano}}, \bibnamefont{and}
  \bibinfo{author}{\bibfnamefont{R.~B.} \bibnamefont{Laughlin}},
  \bibinfo{journal}{Phys. Rev. Lett.} \textbf{\bibinfo{volume}{86}},
  \bibinfo{pages}{3392} (\bibinfo{year}{2001}{\natexlab{a}}).

\bibitem[{\citenamefont{Bernevig
  et~al.}(2001{\natexlab{b}})\citenamefont{Bernevig, Giuliano, and
  Laughlin}}]{Giuliano:2001b}
\bibinfo{author}{\bibfnamefont{B.~A.} \bibnamefont{Bernevig}},
  \bibinfo{author}{\bibfnamefont{D.}~\bibnamefont{Giuliano}}, \bibnamefont{and}
  \bibinfo{author}{\bibfnamefont{R.~B.} \bibnamefont{Laughlin}},
  \bibinfo{journal}{Phys. Rev. Lett.} \textbf{\bibinfo{volume}{87}},
  \bibinfo{pages}{177206} (\bibinfo{year}{2001}{\natexlab{b}}).

\bibitem[{\citenamefont{Bernevig et~al.}(2003)\citenamefont{Bernevig, Giuliano,
  and Santiago}}]{Giuliano:2003}
\bibinfo{author}{\bibfnamefont{B.~A.} \bibnamefont{Bernevig}},
  \bibinfo{author}{\bibfnamefont{D.}~\bibnamefont{Giuliano}}, \bibnamefont{and}
  \bibinfo{author}{\bibfnamefont{D.~I.} \bibnamefont{Santiago}},
  \bibinfo{journal}{Phys. Rev. B} \textbf{\bibinfo{volume}{68}},
  \bibinfo{pages}{115321} (\bibinfo{year}{2003}).

\bibitem[{\citenamefont{von Delft and Schoeller}(1998)}]{Vondelft:1998}
\bibinfo{author}{\bibfnamefont{J.}~\bibnamefont{von Delft}} \bibnamefont{and}
  \bibinfo{author}{\bibfnamefont{H.}~\bibnamefont{Schoeller}},
  \bibinfo{journal}{Ann. Phys.} \textbf{\bibinfo{volume}{7}},
  \bibinfo{pages}{225} (\bibinfo{year}{1998}).

\end{thebibliography}
\end{document}